\documentclass{aa}
\usepackage{txfonts}
\usepackage{natbib}
\usepackage{longtable}
\usepackage{graphicx}
\usepackage{threeparttable}
\bibpunct{(}{)}{;}{a}{}{,} 
	\title{The barium isotopic mixture for the metal-poor subgiant star HD\,140283}
	\author{A. J. Gallagher\inst{1}
\and
S. G. Ryan\inst{1}
\and
A. E. Garc\'ia P\'erez\inst{2,1}
\and
W. Aoki\inst{3}}
\institute{Centre for Astrophysics Research, School of Physics, Astronomy \& Mathematics, University of Hertfordshire, College Lane, Hatfield, Hertfordshire, AL10 9AB, United Kingdom\\ 
email: \texttt{[a.gallagher, s.g.ryan]@herts.ac.uk} 
\and Department of Astronomy, P.O. Box 400325, University of Virginia, Charlottesville, VA 22904-4325, United States\\
email: \texttt{aeg4x@mail.astro.virginia.edu}
\and National Astronomical Observatory, Mitika, Tokyo, 181-8588, Japan\\
email: \texttt{aoki.wako@nao.ac.jp}
}
\date{Received 10 May 2010 / Accepted 20 August 2010}

\authorrunning{A. J. Gallagher et al.}
\titlerunning{barium isotopic mixture for HD\,140283}

\newcommand{\Teff}{T_{\rm eff}}
\newcommand{\kms}{\rm \ km \ s^{-1}}
\newcommand{\fodd}{f_{\rm odd}}
\newcommand{\foddr}{f_{{\rm odd},r}}
\newcommand{\fodds}{f_{{\rm odd},s}}
\newcommand{\foddss}{f_{{\rm odd},ss}}
\newcommand{\logg}{{\rm log} \ g}
\newcommand{\loggf}{{\rm log} \ gf}

\abstract{
Current theory regarding heavy element nucleosynthesis in metal-poor environments states that the \textit{r}-process would be dominant. The star HD\,140283 has been the subject of debate after it appeared in some studies to be dominated by the \textit{s}-process.
}{
We provide an independent measure of the \element{Ba} isotope mixture using an extremely high quality spectrum and an extensive $\chi^2$ analysis.
}{
We have acquired a very high resolution ($R\equiv\lambda/\Delta\lambda = 95\,000$), very high signal-to-noise ($S/N = 1110$ around 4554\,\AA, as calculated in IRAF) spectrum of HD\,140283. We exploit hyperfine splitting of the \ion{Ba}{ii} 4554\,\AA\ and 4934\,\AA\ resonance lines in an effort to constrain the isotope ratio in 1D LTE. Using the code \tiny ATLAS \small in conjunction with \tiny KURUCZ06 \small model atmospheres we analyse 93 \element{Fe} lines to determine the star's macroturbulence. With this information we construct a grid of \element{Ba} synthetic spectra and, using a $\chi^2$ code, fit these to our observed data to determine the isotopic ratio, $\fodd$, which represents the ratio of odd to even isotopes. The odd isotopes and \element[][138]{Ba} are synthesized by the \textit{r}- and \textit{s}-process while the other even isotopes (\element[][134,136]{Ba}) are synthesized purely by the \textit{s}-process. We also analyse the \element{Eu} lines.
}{
We set a new upper limit of the rotation of HD\,140283 at $v\sin{i}\leq3.9\kms$, a new upper limit on ${\rm [Eu/H]} < -2.80$ and abundances ${\rm [Fe/H]} = -2.59\pm 0.09$, ${\rm [Ba/H]} = -3.46\pm0.11$. This leads to a new lower limit on ${\rm [Ba/Eu]} > -0.66$. We find that, in the framework of a 1D LTE analysis, the isotopic ratios of \element{Ba} in HD\,140283 indicate $\fodd=0.02\pm0.06$, a purely \textit{s}-process signature. This implies that observations and analysis do not validate currently accepted theory.
}{
We speculate that a 1D code, due to simplifying assumptions, is not adequate when dealing with observations with high levels of resolution and signal-to-noise because of the turbulent motions associated with a 3D stellar atmosphere. New approaches to analysing isotopic ratios, in particular 3D hydrodynamics, need to be considered when dealing with the levels of detail required to properly determine them. However published 3D results exacerbate the disagreement between theory and observation.
}

\begin{document}
\maketitle

\section{Introduction}
\label{sec:introduction}

Heavy-element abundances are predominantly due to two classes of neutron-capture processes, the rapid (\textit{r}-) process and the slow (\textit{s}-) process. For the \textit{s}-process the beta-decay lifetime is shorter than the timescale for neutron-capture. These two classes can be subdived into the main, weak and strong \textit{s}-process \citep{Clayton67, Busso01, The07, Sneden08} and the main and weak \textit{r}-process \citep{Travaglio04, Wanajo06, Izutani09}. Each neutron-capture process occurs in different environments. The main \textit{s}-process occurs in late-type, low- to intermediate-mass stars ($1 \ {\rm M}_{\sun}\la M \la8 \ {\rm M}_{\sun}$), during thermal pulsing on the asymptotic giant branch (AGB). An uncertain physical event or process is presumed to cause unprocessed \element{H} to mix with \element{C}-rich material in the \element{He}-burning shell to form \element[][13]{C} \citep{Busso01}. In this environment \element[][13]{C} supplies the necessary neutrons via the reaction $\element[][13]{C}(\alpha,{\rm n})\element[][16]{O}$ \citep{BBFH57}. In the core \element{He}-burning phase of solar-metallicity massive stars, where temperatures are relatively high, the nuclear reaction $\element[][22]{Ne}(\alpha,{\rm n})\element[][25]{Mg}$ provides the main source of neutrons, however neutron-capture is mostly weak \textit{s}-process \citep{Pignatari08}. The \element[][22]{Ne} abundance is heavily dependent on the initial CNO abundance and the weak \textit{s}-process produces little \element{Ba} relative to lighter species such as \element{Sr} \citep{Gallino00}.

The astrophysical origin of the \textit{r}-process is still relatively unknown. The most widely proposed site for the \textit{r}-process is when a massive star ($M > 8 \ {\rm M}_{\sun}$) becomes a core-collapse supernova \citep{Wheeler98, Kajino02}. During a core-collapse supernova the neutron flux is believed to be so high that the neutron-capture timescale is shorter than the beta decay lifetime. Other possible \textit{r}-process sites have been considered such as neutron star mergers \citep{FRT99}, however, these seem to have been ruled out as dominant sources for \textit{r}-process material due to their low rates of occurrence \citep{Argast04}. Several theoretical scenarios have been explored in an effort to understand this phenomenon \citep{Wanajo06}. 

The relative importance of the \textit{r}- and \textit{s}-process throughout Galactic history depends on the evolutionary timescales of the proposed sites and their elemental composition. The lifetimes for massive stars are much shorter than for low- to intermediate-mass stars. Typical lifetimes for $25 \ {\rm M}_{\sun}$, $8 \ {\rm M}_{\sun}$, $3 \ {\rm M}_{\sun}$ and $1 \ {\rm M}_{\sun}$ stars are $\sim 7 \times 10^{-3}$, 0.04, 0.4 and 10\,Gyrs respectively \citep{lifetimes05}. As such, the interstellar medium (ISM) at the time at which metal-poor (halo) stars were forming ($\sim$ 12 Gyr ago) should have been enriched by the supernovae of massive stars and hence, \textit{r}-process material. Papers by \citet{SS78} and \citet{Truran81} have been particularly influential in establishing this framework, as we now discuss. 

\citet{SS78} analysed 11 metal-poor halo stars that have a metal abundance less than 1/100 the solar metal abundance. They found that \element{Ba} and \element{Y} were overly deficient relative to \element{Fe}. Both elements can be formed via either neutron-capture process but are dominated by the \textit{s}-process in the solar system where 81\% of \element{Ba} and 92\% of \element{Y} is formed via the \textit{s}-process (Arlandini et al., 1999). The more metal-poor stars in Spite \& Spite's sample had a greater [Ba/Fe]\footnote{[X/Y]=${\rm log}_{10}\big({\frac{N(X)}{N(Y)}}\big)_{*} - {\rm log}_{10}\big({\frac{N(X)}{N(Y)}}\big)_{\sun}$} deficiency than [Y/Fe] deficiency meaning that as [Ba/Fe] decreases, [Ba/Y] decreases also. This is because a greater fluence of neutrons is needed to form \element{Ba} ($Z=56$) than \element{Y} ($Z=39$) so that [Ba/Y] gives a good indication of the number of neutrons captured \citep[see][]{Seeger65}. In contrast, Spite \& Spite found that \element{Eu}, 94\% of which is formed via the \textit{r}-process in solar system material \citep{Arlandini99}, has the same deficiency as \element{Fe}, such that [Eu/Fe] remains constant as [Fe/H] increases and is essentially solar at [Fe/H] $\geq$ -2.6. 

A consideration of the possible sites and seed requirements for neutron-capture led \citet{Truran81} to postulate that neutron-capture-element abundances in metal-poor stars should be dominated by those synthesized through the \textit{r}-process. This expectation arises from the realisation that massive stars are capable of producing both the \element{Fe}-peak seed nuclei and the high neutron fluxes even from very low-metallicity gas, whereas intermediate-mass stars, while capable of producing neutrons, cannot produce the \element{Fe}-peak seeds necessary for the main \textit{s}-process \citep[see also][]{Gallino98}. In Truran's interpretation, the variance of [Y/Fe] and [Ba/Fe] with [Fe/H] are exactly what one would find if the primary source for nuclei beyond the \element{Fe}-peak in metal-poor stars is due to \textit{r}-process nucleosynthesis, while the \textit{s}-process begins to contribute more significantly as [Fe/H] increases giving rise to the increase in [Y/Fe] and [Ba/Fe] seen by \citet{SS78}. He reasoned in addition that the enhancement of \textit{r}-process nuclei would indicate that a prior generation of massive stars formed during or before the formation of the Galaxy. In a more quantitative calculation, \citet{Travaglio99} examined the metallicity dependence of \element{Ba} synthesized in AGB stars via the \textit{s}-process using the chemical evolution model, {\footnotesize FRANEC} \citep{Straniero97,Gallino98}. They found that \element{Ba} formed via the \textit{s}-process has no significant contribution to the \element{Ba} abundance in the Galaxy until [Fe/H] $\ga -1$. This supports Truran's hypothesis.

\begin{figure*}[!htp]
\begin{center}
	\resizebox{!}{0.3\vsize}{\includegraphics{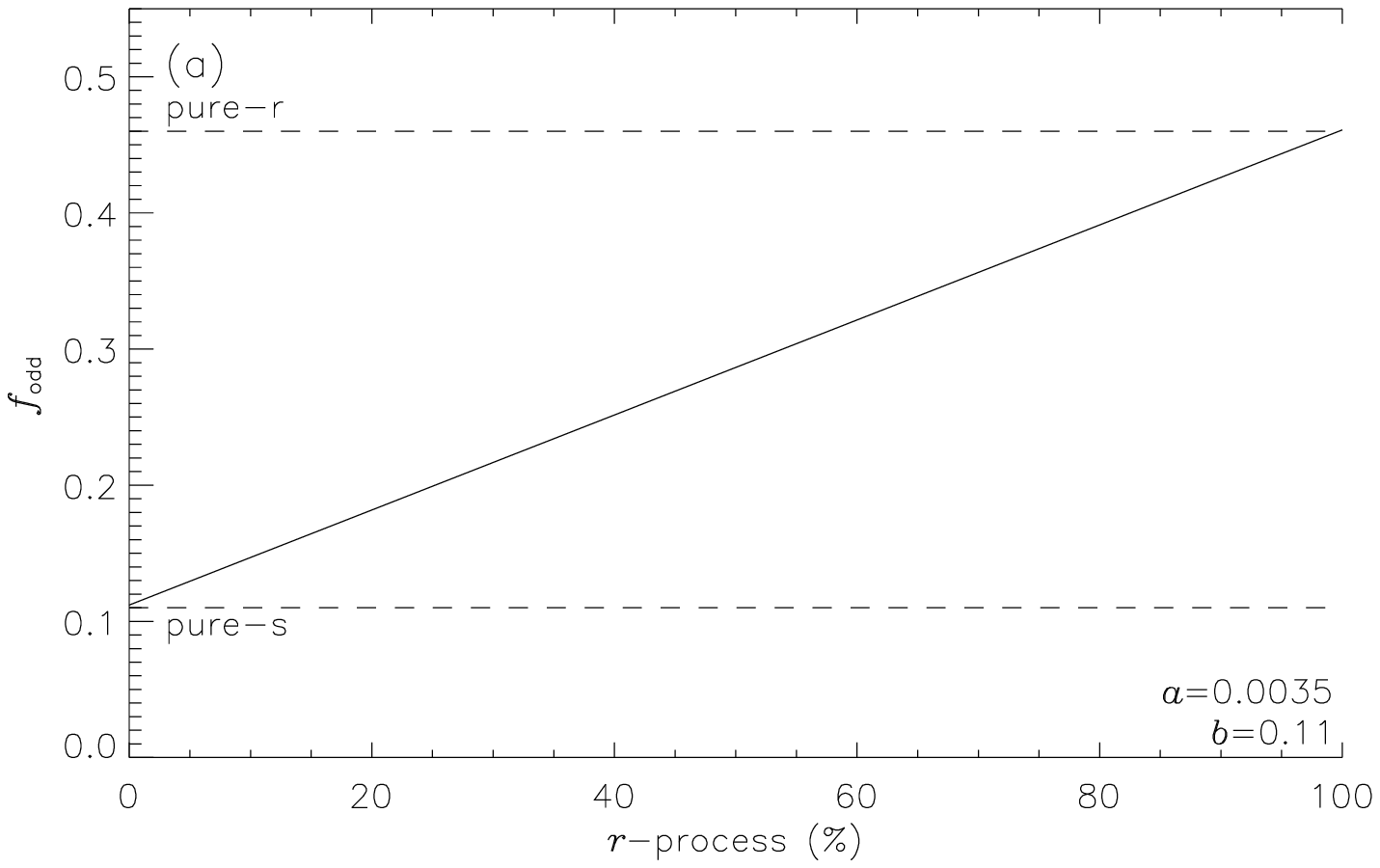}}
	\resizebox{!}{0.3\vsize}{\includegraphics{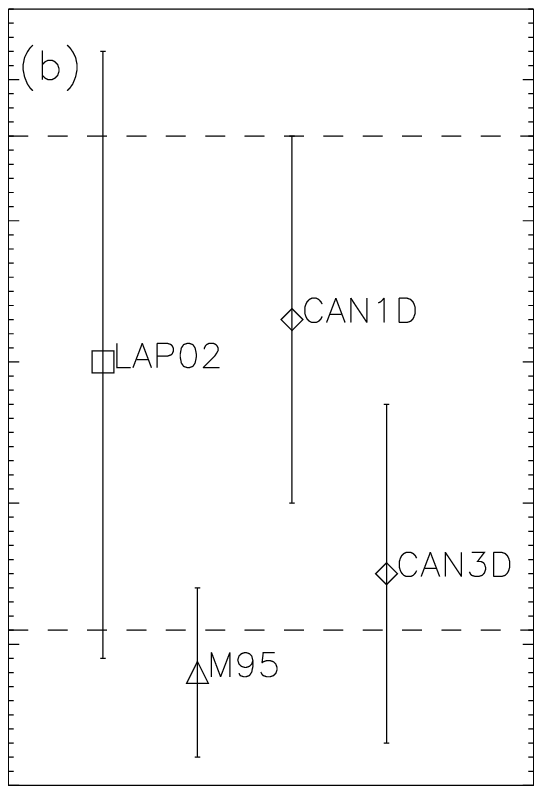}}
	\caption{\bfseries\,a) \mdseries Relation between $\fodd$ and the \textit{r}-process contribution calculated from \citet{Arlandini99}. Coefficients are given where $\fodd$\,=\,\textit{a}\,$\times$\,\textit{r}-process\,(\%)\,+\,\textit{b}. \bfseries\,b) \mdseries \textit{LAP02}: the \citet{Lambert02} result for $\fodd$. \textit{M95}: the \citet{Magain95} result for $\fodd$. \textit{CAN1D}: the \citet{Remo09} 1D LTE result for $\fodd$. \textit{CAN3D}: the \citet{Remo09} 3D hydrodynamical result for $\fodd$.}
	\label{fig:r/fodd}
	\end{center}
\end{figure*}

Different mixtures of odd and even \element{Ba} isotopes are produced by the \textit{r}- and \textit{s}-process. In particular, \element[][134]{Ba} and \element[][136]{Ba} are produced only by the \textit{s}-process due to shielding by \element[][134]{Xe} and \element[][136]{Xe} in the \textit{r}-process. Although the spectral lines of different \element{Ba} isotopes are not well resolved in stellar spectra, the profile core is dominated by the even isotopes while the odd isotopes, which experience hyperfine splitting (hfs), have more importance in the wings of the line profile, relative to the even isotopes. This means that the profiles of \ion{Ba}{ii} 4554\,\AA\ and 4934\,\AA\ are dependent on the contributions of the two processes. Hyperfine splitting arises from the coupling of nuclear spin with the angular momentum of its electrons. The nuclear spin is non-zero in nuclides with odd-\textit{N} and/or odd-\textit{Z}, i.e. where the nucleus has an unpaired nucleon. Hyperfine structure has been well documented in \element{Ba} \citep[e.g.][]{Rutten78, Wendt84, Cowley89, Villemoes93}. As \element{Ba} has an even $Z$, only the two odd isotopes, \element[][135,137]{Ba}, experience hfs. \citet{Arlandini99} calculate that the fraction of odd isotopes\footnote{$\fodd \equiv [N(\element[][135]{Ba}) + N(\element[][137]{Ba})]/N(\element{Ba})$} of \element{Ba} is $\fodds = 0.11\pm0.01$ for a pure \textit{s}-process mixture of \element{Ba} and infer $\foddr = 0.46\pm0.06$ for a pure \textit{r}-process mixture. Their numbers are based on models which best reproduce the main \textit{s}-process using an arithmetic average of 1.5 and 3\,M$_{\sun}$ AGB models at $Z=\frac{1}{2} \ {\rm Z}_{\sun}$. The errors stated here are our propagation of errors associated with individual isotope abundances for \element{Ba}. We show the linear relationship between \textit{r}-process contributions (as a percentage) and $\fodd$ in Fig. \ref{fig:r/fodd}(a).

\citet{Magain95} attempted to verify Truran's proposal by measuring the odd fraction in HD\,140283, a well studied metal-poor subgiant at [Fe/H] = -2.5 \citep{Aoki04}, but found instead that theory and observations were not comparable. He used high-resolution ($R \equiv \lambda/\Delta\lambda = 100\,000$) high signal-to-noise ($S/N \approx 400$) data. Magain reported the fractional odd isotope ratio, $\fodd$, of \element{Ba} to be 0.08$\pm$0.06, implying that \element{Ba} production in HD\,140283 is predominantly due to the \textit{s}-process (see Fig. \ref{fig:r/fodd} (b)) despite [Fe/H] and [Ba/Fe] being very low, [Ba/Fe] = -0.8 \citep{SS78}. The code used to resolve the macroturbulence and analyse the \ion{Ba}{ii} 4554\,\AA\ line solves the equations of hydrostatic equilibrium under the assumption that the stellar atmosphere has a plane-parallel geometry (1D) and local thermodynamic equilibrium (LTE). In contrast some more recent analyses, which we describe below, compute hydrodynamical 3D model atmospheres (3D) where the radiative transfer along multiple lines of sight is assessed.

The star was later reanalysed by \citet{Lambert02}, again assuming 1D LTE. They obtained a very high-resolution ($R \equiv \lambda/\Delta\lambda = 200\,000$) high signal-to-noise ($S/N \approx 550$) spectrum about the \ion{Ba}{ii} 4554\,\AA\ line. They found a value for $\fodd=0.30\pm0.21$ and concluded that, contrary to Magain's result, the star is \textit{r}-process dominated. A value for $\fodd=0.30\pm0.21$ would imply an \textit{r}-process contribution of $54\%\pm60\%$. We note, however, that the error in their measurement of $\fodd$ means that their result covers the full range of possibilities from a pure \textit{s}-process mix ($\fodds = 0.11$) to a pure {r}-process mix ($\foddr = 0.46$); see Fig. \ref{fig:r/fodd}. This means that although they state that their result indicates that HD\,140283 is \textit{r}-process dominated, the range of $\fodd$ is too broad to be conclusive. We consider that an \textit{r}-process contribution of 54\% does not substantially imply that the star's neutron-capture elements are dominated by those synthesised via the \textit{r}-process. 

Against this background, we sought to improve the determination of the \textit{s}-process contribution to \element{Ba} in this star to help us understand the apparent conflicts.

While we conducted our study, \citet{Remo09} reanalysed the \citet{Lambert02} spectrum in 1D LTE and, more significantly, also conducted a new 3D hydrodynamical analysis of the \element{Ba} isotopic fraction of HD\,140283.  In 1D LTE they found that $\fodd=0.33\pm0.13$, meaning that $64\%\pm36\%$ of the \element{Ba} isotopes in HD\,140283 are synthesized via the \textit{r}-process. The central value (0.33) is little changed from that obtained by \citet{Lambert02} (0.30), which is not entirely suprising since they used the same spectrum, but \citet{Remo09} quote smaller error bars. This is because \cite{Lambert02} adopted the standard deviation of macroturbulent broadening estimates as the main underlying measurement error. \citet{Remo09} use the standard error $\sigma/\sqrt{N}$ (where $N$ is the number of \element{Fe} lines used) as a measurement of error. The latter is a more reasonable estimate of the error as it is a measure of the uncertainty in the mean estimate of the broadening. Their analysis of the line using 3D hydrodynamics gives a value for $\fodd = 0.15\pm0.12$ meaning only $11\%\pm34\%$ of the isotopes are synthesized via the \textit{r}-process. This value is in good agreement with the solar \element{Ba} isotopic mix ($\foddss = 0.16$ implying that only 14\% of isotopes formed via the \textit{r}-process \citep{Arlandini99}) but is once more at odds with the high \textit{r}-process fraction expected under Truran's hypothesis.

We have obtained a high resolution ($R \equiv$ $\lambda/\Delta\lambda = 95\,000$) very high signal-to-noise ($S/N = 870 - 1110$) spectrum of HD\,140283. During the course of this paper we discuss how we have constrained the macroturbulence by fitting synthetic spectra to \element{Fe} lines. In a detailed error analysis we show how we have improved constraining the macroturbulence, which was a major source of error that dominated previous studies that analyse the \ion{Ba}{ii} 4554\,\AA\ line in 1D LTE. The improvement is partly due to the higher quality spectrum we have used in this investigation. We also explore the impact of using radial-tangential macroturbulence, $\zeta_{\rm RT}$, and rotational broadening, $v\sin{i}$ (used by \citet{Remo09}) to help constrain macroscopic broadening. We then move on to discuss the method used to re-evaluate the \textit{r}- vs. \textit{s}-process mix by analysing the \ion{Ba}{ii} 4554\,\AA\ line and, for the first time in this context, the \ion{Ba}{ii} 4934\,\AA\ line in 1D LTE. Furthermore we discuss the difficulties in analysing the \ion{Ba}{ii} 4934\,\AA\ line due to close blends with other lines. Also because of the exceptional quality of the data, we have been able to revise downward the \element{Eu} abundance limit for the star.

\section{Observational data}
\label{sec:observation}

Our stellar and \element{ThAr} calibration spectra were obtained over two nights during the commissioning of the High Dispersion Spectrograph (HDS) mounted on the Subaru Telescope. The stellar spectrum is the sum of 13 exposures totalling 82 minutes. This gives a $S/N=1110$ per 12\,m\AA\ wide pixel around the \ion{Ba}{ii} 4554\,\AA\ line and a $S/N=870$ per 12\,m\AA\ wide pixel around the \ion{Ba}{ii} 4934\,\AA\ line, as measured from the scatter in the continuum of the reduced spectrum. The typical resolution as measured from \element{ThAr} lines is $R\equiv\lambda/\Delta\lambda = 95\,000$. The spectrum was reduced using a \element{ThAr} spectrum to wavelength calibrate the stellar spectrum, with typical RMS errors of 1.5\,m\AA\ \citep{Aoki04}. We utilise the 4554\,\AA\ and 4934\,\AA\ lines as both arise from the ground state where hyperfine structure is large. Although the 4934\,\AA\ line is weaker - we measure equivalent widths $W_{4554} = 20.1\,{\rm m}$\AA\ and $W_{4934} = 13.6\,{\rm m}$\AA\ - the hfs of the 4934\,\AA\ line is greater, which means both lines can be useful diagnostics. We do not attempt to analyse higher excitation lines of \element{Ba} which are weaker and have much smaller hfs.

\section{Spectral profiles}
\label{sec:spectralprofile}

To analyse the two \ion{Ba}{ii} line profiles in our spectrum we compared our observed profile to synthetic profiles produced by the 1D LTE code {\footnotesize ATLAS} \citep{Cottrell78}. We describe below how \ion{Fe}{i} and \ion{Fe}{ii} lines are used to constrain macroturbulence, and then proceed to analyse the \ion{Ba}{ii} lines. A 1D {\footnotesize KURUCZ06} model atmosphere ({\ttfamily http://kurucz.harvard.edu/grids.html}) was used with parameters for the star $\Teff = 5750{\rm\ K},\ {\rm [Fe/H]} = -2.5$ and microturbulence, $\xi = 1.4 \kms$ \citep{Aoki04} and $\logg = 3.7 \ [{\rm cgs}]$ \citep{Remo09}.

\subsection{Instrumental Profile}
\label{sec:instrumental}

Two \element{ThAr} hollow-cathode-lamp spectra over the intervals $4102 - 5343$\,\AA\ and $5514 - 6868$\,\AA, taken during the observing run with the same instrumentation and set-up as the stellar exposures used in this study, were used to calculate the instrumental broadening. Using {\footnotesize IRAF}, the full-width at half-maximum (FWHM) and equivalent widths of 993 emission lines were measured. It was found that at a wavelength of 4554\,\AA\ the \element{ThAr} line FWHM in velocity space ($\nu_{\rm inst}$) was $3.31\kms$, and at 4934\,\AA\ was $3.25\kms$. The error in these measurements is taken as the standard error of the mean of the individual measurements, $\sigma/ \sqrt{N}$, which is $0.01\kms$, where $\sigma$ is the standard deviation of the individual measurements, which is $0.22 \kms$. We assume here that the \element{ThAr} lines are unresolved and hence that the measured \element{ThAr} line width represents the instrumental broadening. The instrumental broadening could be slightly less than that stated, but the difference is immaterial since, in \S\ref{sec:macroturbulence}, we measure the combined instrumental and macroturbulent broadening without needing to distinguish between the two contributions precisely. \citet{Aoki04} showed that the instrumental profile is well approximated by a Gaussian.

\subsection{Macroturbulence}
\label{sec:macroturbulence}
\citet{Lambert02} established that one of the major limiting factors in their analysis was the accuracy with which macroturbulence could be measured. They found that $\delta \fodd / \delta {\rm FWHM} = -0.51 \ ({\rm km \ s^{-1}})^{-1}$, and hence for $\sigma_{\rm FWHM} = 0.33 \kms$ they achieved an accuracy in macroturbulence corresponding to $\sigma_{\fodd} \sim 0.17$, dominating their total error of 0.21. It was clear that we would have to improve on this significantly to make progress. 

\begin{figure}[!h]
	\begin{center}
	\resizebox{\hsize}{!}{\includegraphics{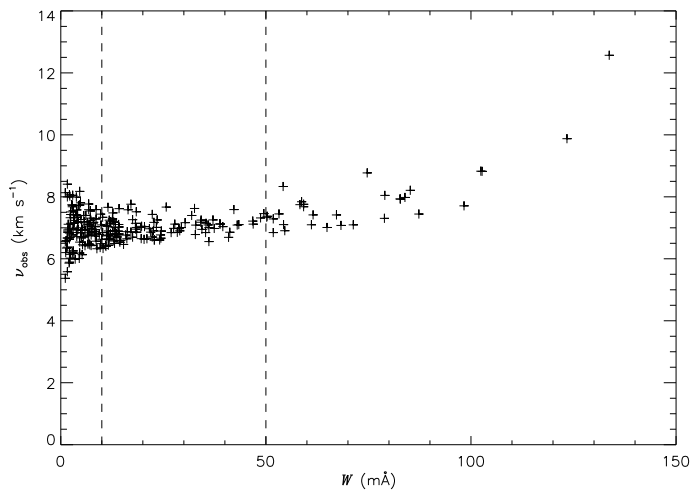}}
	\caption{FWHM versus equivalent width for the 257 \ion{Fe}{i} and \ion{Fe}{ii} measured for HD\,140283. 93 lines fall into the range of $10 \leq W$ (m\AA) $\leq 50$; the mean value for $\nu_{\rm obs}$ in this range is $6.9 \kms$.}
	\label{fig:Vfwhm}
	\end{center}
\end{figure}

We began to constrain macroturbulence by measuring the equivalent widths and the FWHM (in velocity space), $\nu_{\rm obs}$, of 257 apparently unblended \ion{Fe}{i} and \ion{Fe}{ii} lines by fitting Gaussian profiles in {\footnotesize IRAF}. We used this information to produce Fig. \ref{fig:Vfwhm}. As $\Delta\lambda$/$\lambda$ remains constant with wavelength in an echelle spectrum (where $\Delta\lambda$ is the width of the pixel in wavelength units) it is possible to use Fig. \ref{fig:Vfwhm} as a check of the quality of measurements.  

From Fig. \ref{fig:Vfwhm} we can see that weaker lines, $W \leq 50\,{\rm m}$\AA, almost form a plateau. Here, $\nu_{\rm obs}$ remains constant even as the Doppler core deepens in lines on the linear part of the curve of growth, where the Doppler broadening components are dominant. At $W > 50 \ {\rm m}$\AA, pressure broadening become significant as the core of the line saturates, so the wings begin to broaden. Where $W < 10 \ {\rm m}$\AA, the uncertainty produced by the finite signal-to-noise makes it difficult to measure the lines accurately, which is shown by the scatter in this region of Fig. \ref{fig:Vfwhm}.

Of the 257 \element{Fe} lines measured, 93 fell between 10\,m\AA\,$\leq W \leq$\,50\,m\AA\ and were used to constrain macroturbulence (recall that $W_{4554} = 20.1$\,m\AA\ and $W_{4934} = 13.6$ m\AA). The average value for the observed velocity FWHM, $\nu_{\rm obs}$, in this range is $6.9 \kms$. The full list of measurements can be found in Table \ref{tab:Felines}. 

\subsubsection{Gaussian macroturbulence}
\label{sec:gaussian}

We convolve the synthetic flux spectrum of the star with a Gaussian of FWHM $\nu_{\rm conv}$ which represents the convolution of the Gaussian instrumental profile with a Gaussian macroturbulent profile. For now we assume that the star has no significant rotation; we shall return to this point in \S \ref{sec:rotation}. Current estimates of rotation of HD\,140283 are $v\sin{i} = 5.0\pm2.0\kms $ \citep{Medeiros06}. We  create a grid of 385 convolved synthetic spectra for 11 values of macroturbulence $4.9 \kms \leq \nu_{\rm conv} \leq 6.9 \kms$ in steps $\Delta\nu_{\rm conv}=0.1\kms$ and 35 values for $A$(\element{Fe})\footnote{$A({\rm X}) \equiv {\rm log}_{10}\big(\frac{N({\rm X})}{N({\rm H})}\big) + 12$}, $4.09 \leq A({\rm \element{Fe}}) \leq 5.45$ with steps $\Delta A({\rm \element{Fe}}) = 0.04$. Each synthetic spectrum covered the wavelength range $4100 - 6900$\,\AA\ in intervals of $\Delta\lambda=0.01$\,\AA.

\begin{figure}[!ht]
	\begin{center}
	\resizebox{\hsize}{!}{\includegraphics{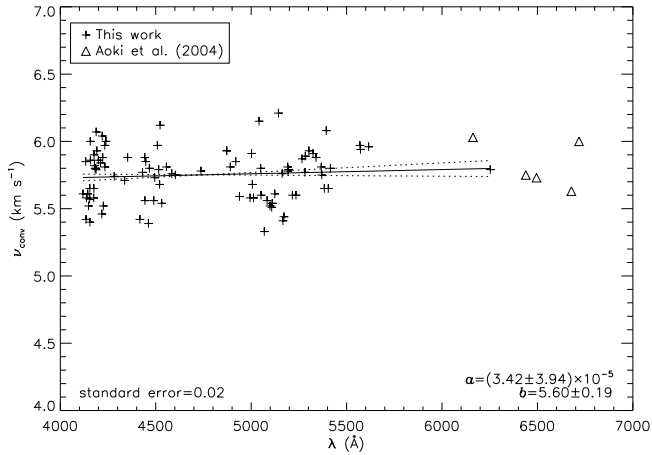}}
	\caption{Values of $\nu_{\rm conv}$ that satisfy the minimum value for $\chi^{2}$ for the 93 lines (\textit{plus symbols}). The standard error represents the scatter from the mean of each line ($\sigma/\sqrt{N}$) and $a$ and $b$ are coefficients of the least squared fit, $\nu_{\rm conv} = a\lambda + b$. The \ion{Ca}{i} and \ion{Fe}{i} lines used by \citet{Aoki04} to constrain macroturbulence have been plotted (\textit{triangles}) to show the consistency of our $\chi^{2}$ fits with their results at red wavelengths.}
	\label{fig:macro}
	\end{center}
\end{figure}

To determine the best fit for $\nu_{\rm conv}$ we compare our synthetic model grid to the observed spectrum employing a $\chi^{2}$ test, $\chi^{2} \equiv \sum(O_i-M_i)^{2}/{\sigma^{2}_{i}}$, where $O_i$ is the observed continuum-normalised profile, $M_i$ is the model profile of the line produced using {\footnotesize ATLAS} and $\sigma_{i}^{2}$ is the standard deviation of the observed points that define the continuum, i.e. $\sigma = (S/N)^{-1}$. All 93 \ion{Fe}{i} and \ion{Fe}{ii} lines were individually fitted using a $\chi^2$ code \citep{Ana09}. This code allows small wavelength shifts, $\Delta \lambda$, which we discuss below and local renormalisation of the continuum of the observed profile for every line. It finds values for $\Delta \lambda$, $A$(\element{Fe}) and macroturbulence that minimize $\chi^2$ for each \element{Fe} line analysed over a window 0.6\,\AA\ wide and with continuum windows typically 0.5\,\AA\ to 1.0\,\AA\ on each side of this, depending on neighbouring spectral features. Values of $\nu_{\rm conv}$ found by the $\chi^2$ code for the 93 lines covering the wavelength range $4118-6253$\,\AA\ are shown in Fig. \ref{fig:macro}. The full table of results from the $\chi^2$ code is found in Table \ref{tab:Felines}.

We use an ordinary least squares (OLS) fit to determine $\nu_{\rm conv}$ at the wavelengths 4554\,\AA\ and 4934\,\AA. The OLS has the equation $\nu_{\rm conv} = a\lambda + b$, where \textit{a} and \textit{b} are coefficients of the OLS. We find that $\nu_{\rm conv}=5.75 \kms$ and $\nu_{\rm conv}=5.76 \kms$ at the \ion{Ba}{ii} 4554\,\AA\ and 4934\,\AA\ lines respectively. The error in these values, represented by the standard error, is $0.02 \kms$. As the uncertainty in these values is greater than the difference between them, we adopted one value for $\nu_{\rm conv}$ for both \ion{Ba}{ii} lines, $\nu_{\rm conv} = 5.75 \pm 0.02 \kms$. Subtracting the instrumental FWHM at 4554\,\AA\ we find the macroturbulence, $\nu_{\Gamma}$, to be $\nu_{\Gamma} = \sqrt{5.75^2 - 3.31^2} = 4.70 \pm 0.02 \kms$. This value agrees well with that found by \citet{Aoki04}. The error in $\nu_{\Gamma}$ is given by $\sigma_{\nu_{\Gamma}}^2 = \sigma_{\nu_{\rm conv}}^2 + \sigma_{\nu_{\rm inst}}^2$, which is equal to $\pm 0.02\kms$. 

In using the \element{Fe} lines to determine the macroturbulence appropriate to \element{Ba}, it is important that we measure lines forming over a similar range of depths in the photosphere. This was achieved in the first instance by restricting the equivalent width range of the \element{Fe} lines to span the two \element{Ba} lines (see Fig \ref{fig:Vfwhm}). In addition, we have regressed the $\nu_{\rm conv}$ measurements against equivalent width, $W$, and against excitation energy, $\chi$, and find no statistically significant trend of $\nu_{\rm conv}$ with $W$, and a weak ($2.5\sigma$) trend with $\chi$. This suggests that using a stricter restriction on the \element{Fe} line list would not materially alter the macroturbulent velocity. In the most extreme case, the value for $\chi=0$\,eV would imply $\nu_{\rm conv} = 5.62\pm0.05\kms$, which (as we show below) would increase $\fodd$ by 0.09.

The $\chi^2$ code also determined that $A (\element{Fe}) =  4.91 \pm 0.01$, where the error is taken as the standard error. Taking the solar \element{Fe} abundance to be $A({\rm Fe})_{\sun} = 7.50 \pm 0.05$ from \citet{Grevesse98}, we calculate the metallicity, ${\rm [Fe/H]} = -2.59\pm0.05$, where the error in [Fe/H] is the propagation of the statistical error in $A(\element{Fe})_*$ and $A(\element{Fe})_{\sun}$ but so far excludes the systematic errors associated with the imperfect choice of atmospheric parameters. That error, based on calculations we provide in \S \ref{sec:uncertainty}, is around 0.07 dex, giving a total error of 0.09 dex. This is in good agreement with metallicity we adopted from \citet{Aoki04}. We note that there is an updated list of solar abundances given in \citet{Grevesse07} calculated using 3D hydrodynamics, however we decided to use the 1D LTE results given in \citet{Grevesse98} as we are working in 1D LTE.

We found that the mean wavelength shift, $\Delta \lambda$, was $-12.0$\,m\AA\ with a standard deviation $\sigma_{\Delta\lambda}=3.8 \ {\rm m}$\AA. There are several reasons why we would expect to find a wavelength shift between the observed and synthetic profiles. The most likely is an error in the approximate radial velocity correction of the star, but line-to-line differences require further comment. There could be inaccuracies in the assumed wavelengths in the \element{Fe} line list, however the \element{Fe} line list was produced using the most up to date data available through the {\footnotesize IRON PROJECT} and \citet{Nave94}, where wavelengths are quoted to 1\,m\AA, and are believed to be accurate to $<1$\,m\AA. The RMS error in the wavelength calibration was reported as only 1.5\,m\AA\ \citep{Aoki04}, so the line-to-line scatter $\sigma_{\Delta\lambda}$ exceeds that error. The excess could be due to the inability of 1D hydrostatic model atmospheres to model turbulent motions in a star's hydrodynamic atmosphere. Indeed, the residuals shifts were found to depend, at least partially, on the excitation potential, $\chi$, and the equivalent width, $W$, suggesting an astrophysical cause. 

\subsubsection{Non-Gaussian symmetric broadening}
\label{sec:rotation}

So far we have adopted a Gaussian macroturbulent broadening mechanism. We looked at two other macro-scale broadening mechanisms, radial-tangential macroturbulence ($\zeta_{\rm rt}$) and rotation ($v\sin{i}$) \citep[][Chapter 18]{Graybook}. Each broadening run was given the same atmospheric parameter set; $\Teff = 5750 \ {\rm K}$, $\logg=3.7$, ${\rm [Fe/H]} = -2.5$, $\xi = 1.4 \kms$. Table \ref{tab:broad} shows the results from fitting the 93 \element{Fe} lines using the three broadening types, along with Gaussian instrumental broadening ($\nu_{\rm inst} = 3.31 \kms$). The third column indicates how many of the 93 lines were best fit by that broadening mechanism, as judged by the minimum $\chi^2$ value for the three methods.

\begin{table}[!ht]
\begin{center}
\caption{Comparison of all three broadening types. Column two gives the broadening based on all 93 \element{Fe}  lines for the wavelength 4554\,\AA, determined using the method described in \S \ref{sec:gaussian}. The errors given are the standard error ($\sigma/\sqrt{N}$). Column three shows how many of the \element{Fe}  lines were statistically better fits with that particular broadening technique.}
\begin{tabular}{ccc}
\hline\hline
Broadening Parameter 		 & Parameter ($\kms$) & \# of best fit lines \\ 
\hline
$\nu_\Gamma$  			 & 						$4.70\pm0.02$		    &		32	 	    \\		
$\zeta_{\rm rt}$  	 &						$4.37\pm0.02$				&		58		    \\
$v\sin{i}$  					 &						$3.89\pm0.02$				&		3			    \\
\hline
\end{tabular}
\label{tab:broad}
\end{center}
\end{table}	

Only three of the 93 \element{Fe} lines were fit best by rotational broadening, and hence we concluded that using only rotational velocities to broaden the lines would be unsound. The derived value provides a firm upper limit on rotation, $v\sin{i}\leq3.9\kms$, in the case with no macroturbulent broadening. The fact that most lines are fit better by a macroturbulent profile emphasises that we have not detected true rotation of the star at this $3.9\kms$ level. 

The radial-tangential broadening function within the 1D LTE framework, provides a better fit than Gaussian macroturbulence to almost two thirds of the \element{Fe} lines. We present \element{Ba} results for both macroturbulent prescriptions in \S \ref{sec:r-process}, but unless specified, our analysis is conducted using Gaussian fitting.  

\section{The \ion{Ba}{ii} resonance lines and the barium isotopic ratio}
\label{sec:bariumlines}

\subsection{\ion{Ba}{ii} line structure}
\label{sec:bariumstructure}

There are five principal, stable \element{Ba} isotopes that are formed via the two neutron-capture processes. The \textit{r}- \& \textit{s}-process produce different mixes of odd and even isotopes. The \textit{r}-process does not contribute to two even isotopes, \element[][134,136]{Ba}, which are pure \textit{s}-process isotopes. The two odd isotopes, \element[][135,137]{Ba}, and even isotope \element[][138]{Ba} are formed from both the \textit{s}- \& \textit{r}-process. The odd isotopes broaden the line and make it asymmetric, whereas the even isotopes contribute to the centre of the \ion{Ba}{ii} line and make the core deeper. 

\begin{table}[!ht]																											
\begin{center}																											
\caption{The isotopic and hfs information for both \element{Ba} lines. The	oscillator	strengths	relative	to	\element[][138]{Ba}	for	each	line	are	given	in	column	3 and the calculated \textit{gf}-values are given in columns 4 and 5.}
\begin{tabular}{c	c	c	c	c	}																						
\hline\hline																											
	&	&	&	\multicolumn{2}{c}	{\textit{gf}-value}	\\																					
\cline{4-5}																											
$\lambda$	(\AA)	&	Isotope	&	Relative	strength	&	\textit{s}-process	&	\textit{r}-process	\\																
\hline
\vspace{-3mm} \\																											
4553.9980	&	137	&	0.1562	&	0.0210	&	0.0471	\\																		
4553.9985	&	137	&	0.1562	&	0.0210	&	0.0471	\\																		
4553.9985	&	137	&	0.0626	&	0.0084	&	0.0189	\\																		
4554.0010	&	135	&	0.1562	&	0.0049	&	0.0594	\\																		
4554.0015	&	135	&	0.1562	&	0.0049	&	0.0594	\\																		
4554.0020	&	135	&	0.0626	&	0.0019	&	0.0238	\\																		
4554.0316	&	134	&	1.0000	&	0.0429	&	0.0000	\\																		
4554.0319	&	136	&	1.0000	&	0.1450	&	0.0000	\\																		
4554.0332	&	138	&	1.0000	&	1.1256	&	0.7972	\\																		
4554.0474	&	135	&	0.4376	&	0.0136	&	0.1663	\\																		
4554.0498	&	137	&	0.4376	&	0.0589	&	0.1320	\\																		
4554.0503	&	135	&	0.1562	&	0.0049	&	0.0594	\\																		
4554.0513	&	135	&	0.0311	&	0.0010	&	0.0118	\\																		
4554.0537	&	137	&	0.1562	&	0.0210	&	0.0471	\\																		
4554.0542	&	137	&	0.0311	&	0.0042	&	0.0094	\\																		
\vspace{-2mm}\\																											
4934.0288	&	137	&	0.3125	&	0.0197	&	0.0441	\\																		
4934.0332	&	135	&	0.3125	&	0.0045	&	0.0556	\\																		
4934.0410	&	137	&	0.0625	&	0.0039	&	0.0088	\\																		
4934.0439	&	135	&	0.0625	&	0.0009	&	0.0111	\\																		
4934.0750	&	134	&	1.0000	&	0.0201	&	0.0000	\\																		
4934.0753	&	136	&	1.0000	&	0.0678	&	0.0000	\\																		
4934.0768	&	138	&	1.0000	&	0.5265	&	0.3729	\\																		
4934.0918	&	135	&	0.3125	&	0.0045	&	0.0556	\\																		
4934.0942	&	137	&	0.3125	&	0.0197	&	0.0441	\\																		
4934.1025	&	135	&	0.3125	&	0.0045	&	0.0556	\\																		
4934.1064	&	137	&	0.3125	&	0.0197	&	0.0441	\\																		
\hline																											
\label{tab:linelist}																											
\end{tabular}																											
\end{center}																											
\end{table}																											

\begin{figure}[!ht]
	\begin{center}
	\resizebox{\hsize}{!}{\includegraphics{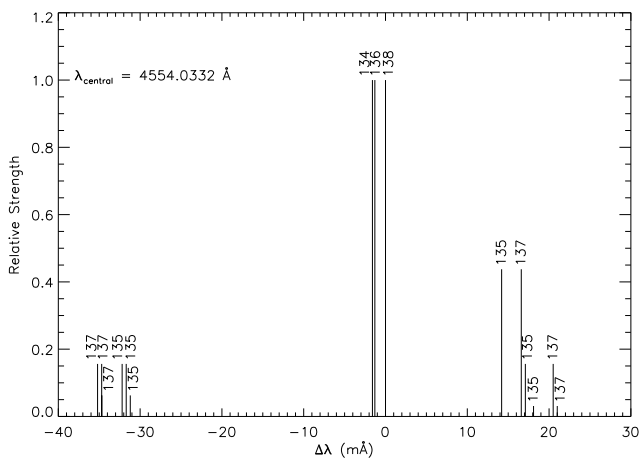}}
	\caption{The splitting patterns of the \ion{Ba}{ii} 4554\,\AA\ line relative to \element[][138]{Ba}. The relative strengths of each isotope are normalised to 1 (Table \ref{tab:linelist}, column (3)).}
	\label{fig:Split}
	\end{center}
\end{figure}

We use the hfs information from \citet{Wendt84} and \citet{Villemoes93} to compute energy level splittings for the lower and upper levels of \element[][135,137]{Ba} in the 4554\,\AA\ and 4934\,\AA\ lines, and incorporated the isotopic shifts relative to \element[][138]{Ba}. Importantly, we note that the hfs splitting and isotope shift data in these papers is quite similar to that of some older studies \citep[references can be found in][]{Wendt84,Villemoes93}, and hence we are confident that there is no significant uncertainty in the line structure. The line wavelength structure, relative to \element[][138]{Ba}, is shown in Fig. \ref{fig:Split}, in which each isotope is shown with a total strength of 1.0. Using the \textit{s}- and \textit{r}-process contributions to the five isotopes from \citet{Arlandini99}, we construct line lists for pure \textit{s}-process and pure \textit{r}-process isotope mixes for the two \ion{Ba}{ii} lines, adopting total $\loggf$ values of +0.16 and -0.16 for 4554\,\AA\ and 4934\,\AA\ respectively, see Table \ref{tab:linelist}. From these two lists we created a further 13 hybrid line lists. These covered a range for $\fodd$ equal to $0.00 \leq \fodd \leq 0.46$ where $\Delta \fodd=0.035$. We recognise that according to \citet{Arlandini99}, the cases with $\fodd<0.11$ are not achieved astrophysically because even a pure \textit{s}-process mixture has a non-zero contribution of \element[][135,137]{Ba}.

\subsection{$\chi^2$ test}
\label{sec:chi^2}

The observed continuum was renormalised over a window of 1\,\AA\ either side of each of the two \element{Ba} lines. A new grid comprising 231 synthetic spectra around each of the two \ion{Ba}{ii} resonance lines was produced in {\footnotesize ATLAS}. Values for $\nu_{\rm conv}$ and \textit{A}(Fe), constrained in the last section, were fixed. There were three free parameters in the new grid: \textit{A}(Ba), $\Delta\lambda$ and the \textit{r}- \& \textit{s}-process contributions. The $\chi^2$ code allowed small changes in these parameters exactly like the code described in \S \ref{sec:gaussian}. We used 21 values for \textit{A}(\element{Ba}), $-1.40 \leq A({\rm Ba}) \leq -1.20$, where $\Delta A{\rm (\element{Ba})}=0.01$. Each synthetic spectrum covered the range $4550 - 4560$\,\AA\ (around 4554\,\AA), $4930 - 4940$\,\AA\ (around 4934\,\AA) and was computed every 0.01\,\AA. The windows in which both \element{Ba} lines were analysed was $\pm0.25$\,\AA\ from their centroid. 

\subsection{The iron blends at 4934\,\AA.}
\label{sec:ironblend}

It has been documented that the \ion{Ba}{ii} 4934\,\AA\ line has a known blend with a weak \ion{Fe}{i} line \citep{Cowley89}. We use the information for two \element{Fe} lines which are found in \citet[their table 2]{Nave94}. The relevant data can be found in our Table \ref{tab:feblend}.

\begin{figure}[!ht]
	\begin{center}
	\resizebox{\hsize}{!}{\includegraphics{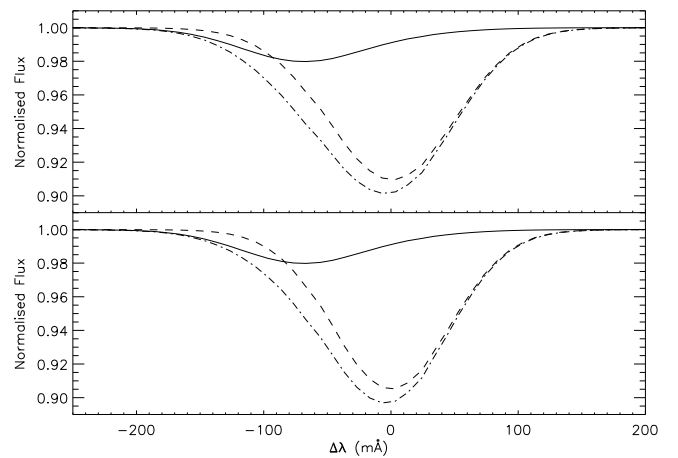}}
	\caption{Synthetic spectra showing the effect of the \ion{Fe}{i} blends on the 4934\,\AA\ \ion{Ba}{ii} line. The top plot shows the \textit{r}-process-only isotope fraction and the bottom shows the \textit{s}-process-only isotope fraction. \textit{Solid line}: the underlying \ion{Fe}{i} blends. \textit{Dashed line}: the uncontaminated \ion{Ba}{ii} line. \textit{Dash-dot line}: the overall line profile.}
	\label{fig:blend}
	\end{center}
\end{figure}

\begin{table}[!ht]
\caption{Spectroscopic information on the two weak \element{Fe} lines that are blended with the \ion{Ba}{ii} 4934\,\AA\ line as reported by \citet{Nave94}. The $\loggf$ values reported here were supplied by G. Nave (2009, private communication) but can be found using the Kurucz database {\tiny \texttt{http://www.cfa.harvard.edu/amp/ampdata/kurucz23/sekur.html}}.}
\begin{center}
\begin{tabular}{c c c c}
\hline\hline
$\lambda$\,(\AA) & Ion   			 & $\chi$ (eV) & $\loggf$ \\
\hline
4934.0052 			 & \ion{Fe}{i} & 4.15	 		   &  -0.589	\\
4934.0839				 & \ion{Fe}{i} & 3.30	       &	-2.307	\\ 
\hline
\label{tab:feblend}
\end{tabular}
\end{center}
\end{table}

When these wavelengths are compared with the wavelengths for the \element{Ba} 4934\,\AA\ line given in Table \ref{tab:linelist} it is clear that the two \element{Fe} lines would influence the \textit{r}-process fraction found by analysis of the line. This is shown in Fig. \ref{fig:blend} for pure \textit{s}- and \textit{r}-process isotope ratios. The analysis of the 4934\,\AA\ line is very sensitive to the characteristics of the \element{Fe} lines, as we show in \S \ref{sec:4934uncert}. We include these \element{Fe} lines in our \element{Ba} 4934\,\AA\ line analysis.

\subsection{The \textit{r}-process contribution}
\label{sec:r-process}

The \element{Ba} abundances of the two lines are found to be $A(\element{Ba})_{4554} = -1.28$ and $A(\element{Ba})_{4934} = -1.30$. We take the implied \element{Ba} abundance as an average of the two, $A(\element{Ba}) = -1.29\pm0.08$. Using the solar abundances calculated in \citet{Grevesse98} we find that for HD\,140283, ${\rm [Ba/H]} = -3.46\pm0.11$, and hence ${\rm [Ba/Fe]} = -0.87\pm0.14$. Errors stated here are calculated in \S \ref{sec:uncertainty}. Results from other papers are given in Table \ref{tab:result}. It is shown that our result for [Ba/Fe] is in good agreement with previous results.

Our $\chi^2$ fitting procedure included a possible wavelength shift as a free parameter. The 4554\,\AA\ line has a wavelength shift $\Delta\lambda = -14.8$\,m\AA. The 4934\,\AA\ line has a wavelength shift $\Delta\lambda = -21.5$\,m\AA, possibly because of imperfect modelling of the \element{Fe} blend in the blue wing. Both lines fall within $3\sigma$ of the mean wavelength shift found in the \element{Fe} lines, -12.0\,m\AA\ with $\sigma_{\Delta\lambda}=3.8$\,m\AA.

From the $\chi^2$ analysis we find the best statistical fit for the 4554\,\AA\ line is $\fodd=0.01\pm0.06$. The best statistical fit for the 4934\,\AA\ line indicates a value of $\fodd = 0.11\pm0.19$ (meaning an \textit{r}-process contribution of $0\%\pm54\%$). The $1\sigma$ errors stated here arise from uncertainties discussed in \S \ref{sec:uncertainty} and are larger for the 4934\,\AA\ line because of uncertainties associated with the underlying \element{Fe} blends. Our result is in good agreement with \citet{Magain95} who found that \ion{Ba}{ii} 4554\,\AA\ yielded $\fodd = 0.08\pm0.06$, but seems to be at odds with values found by \citet{Lambert02} ($\fodd = 0.30\pm0.21$) and the 1D result found for the same spectrum by \citet{Remo09} ($\fodd = 0.33\pm0.13$).

\begin{figure*}[!ht]
	\begin{center}
	\resizebox{0.49\hsize}{!}{\includegraphics{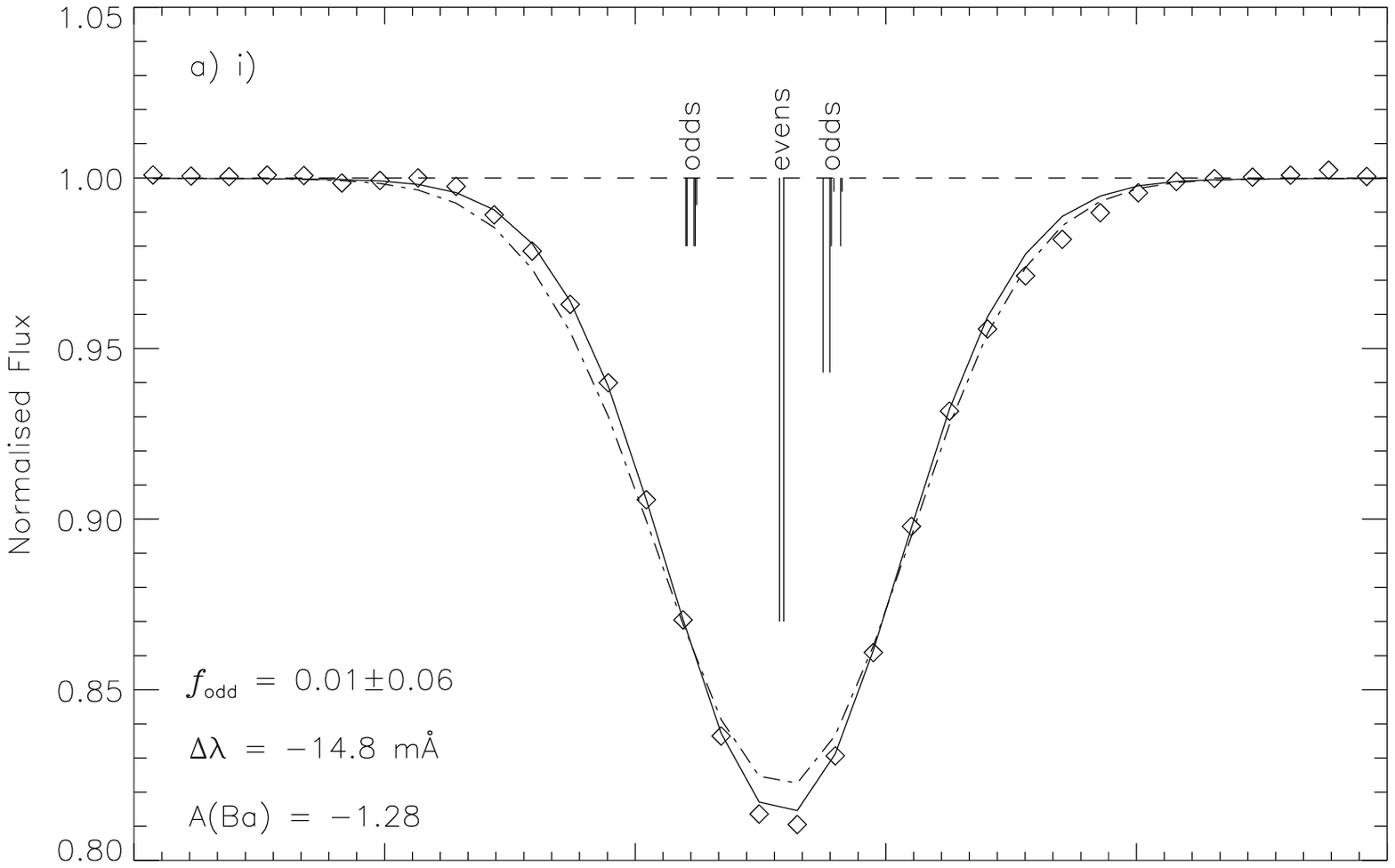}}
	\resizebox{0.49\hsize}{!}{\includegraphics{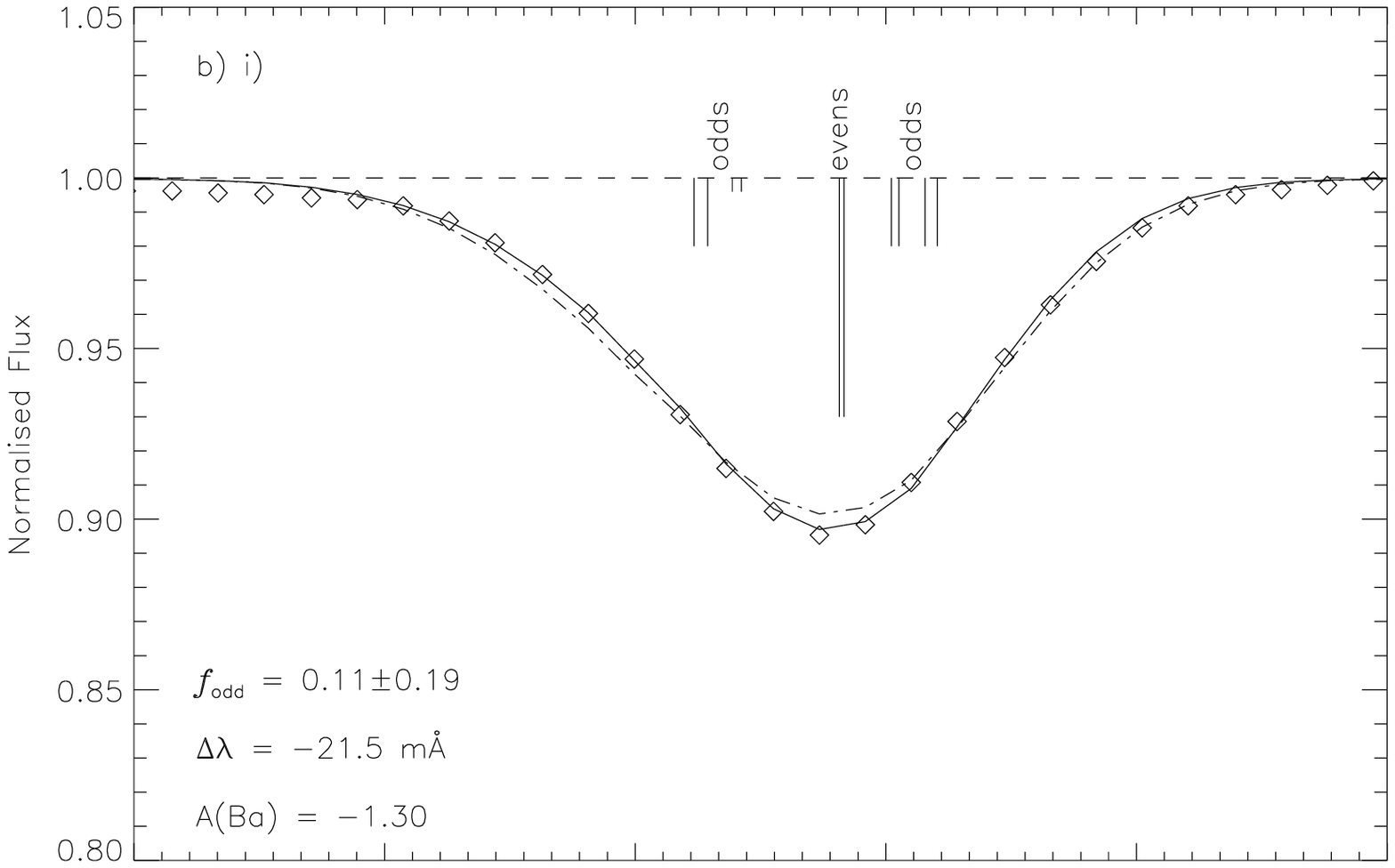}}
	\resizebox{0.49\hsize}{!}{\includegraphics{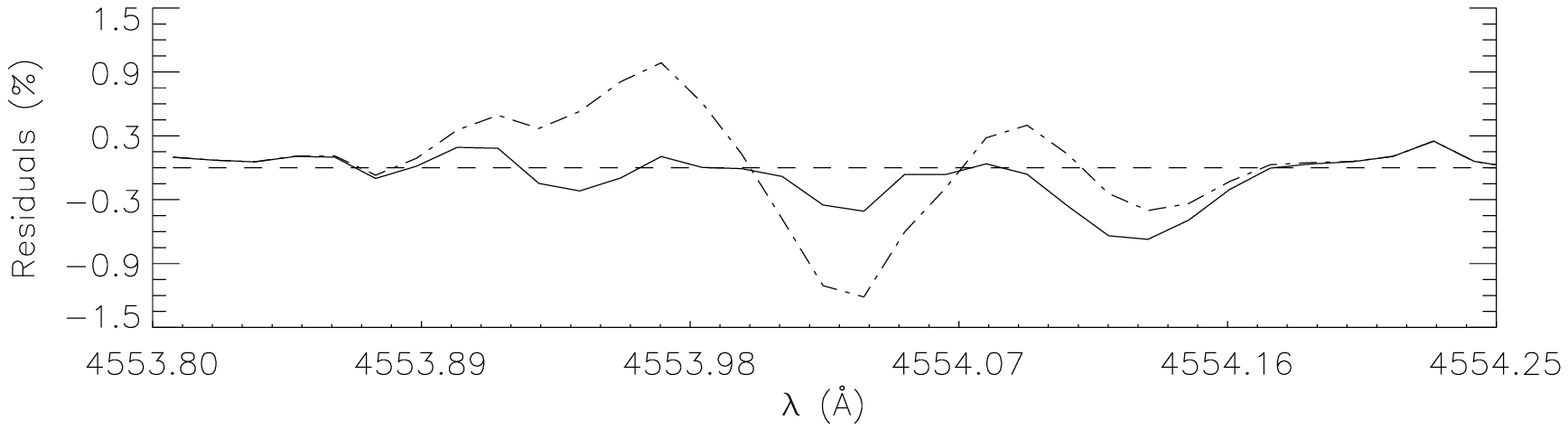}}
	\resizebox{0.49\hsize}{!}{\includegraphics{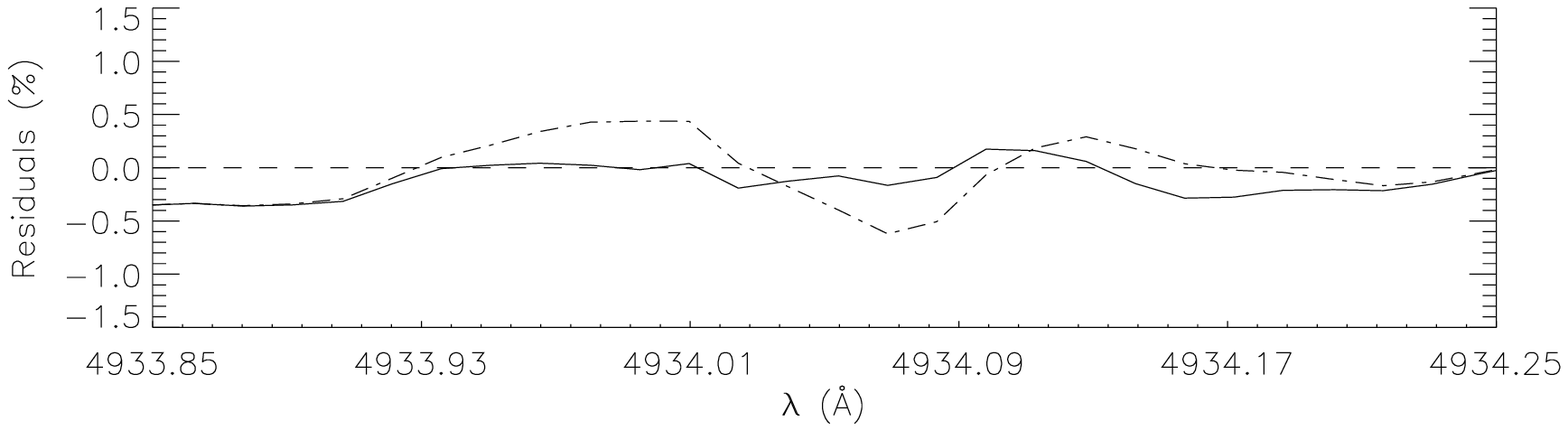}}
	\resizebox{0.49\hsize}{!}{\includegraphics{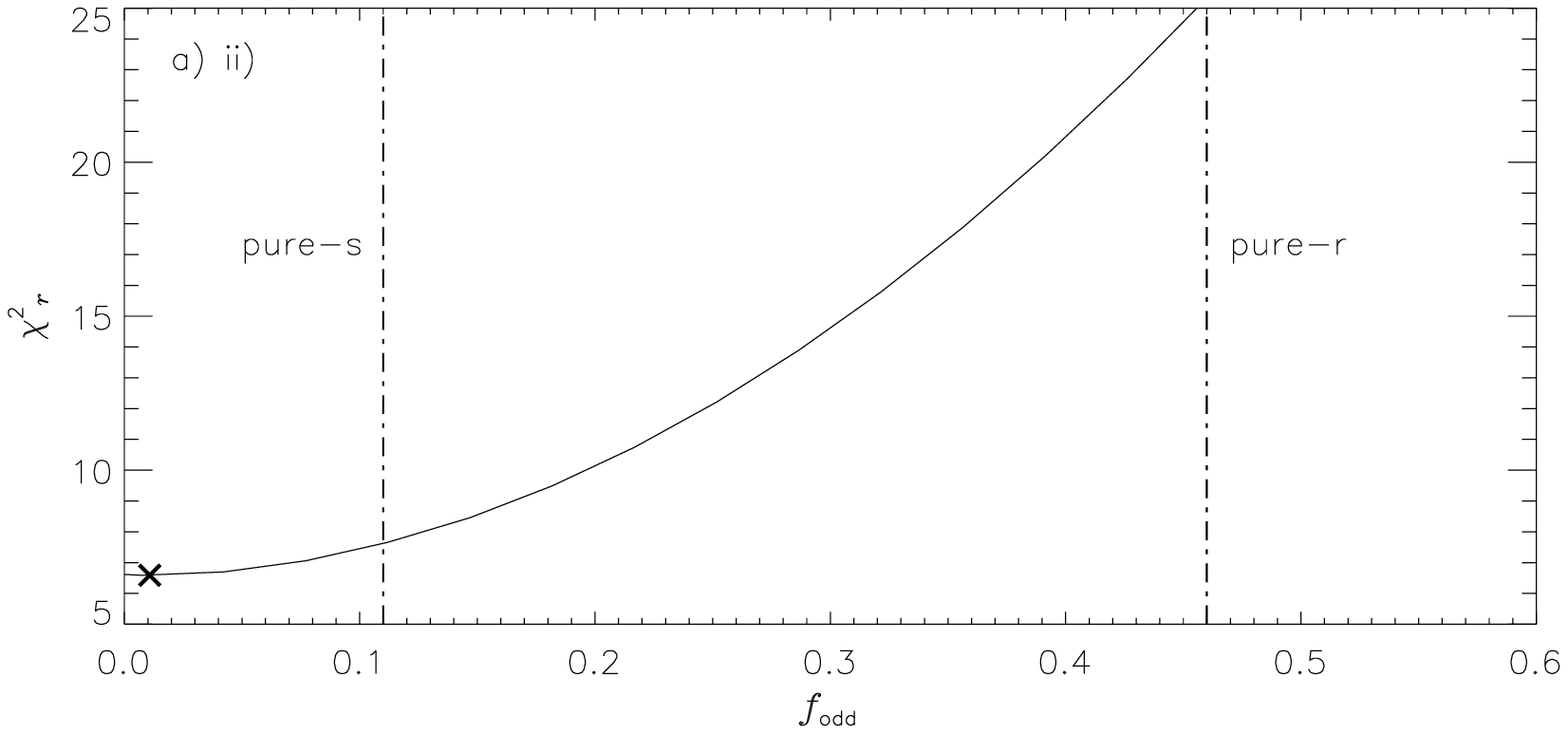}}
	\resizebox{0.49\hsize}{!}{\includegraphics{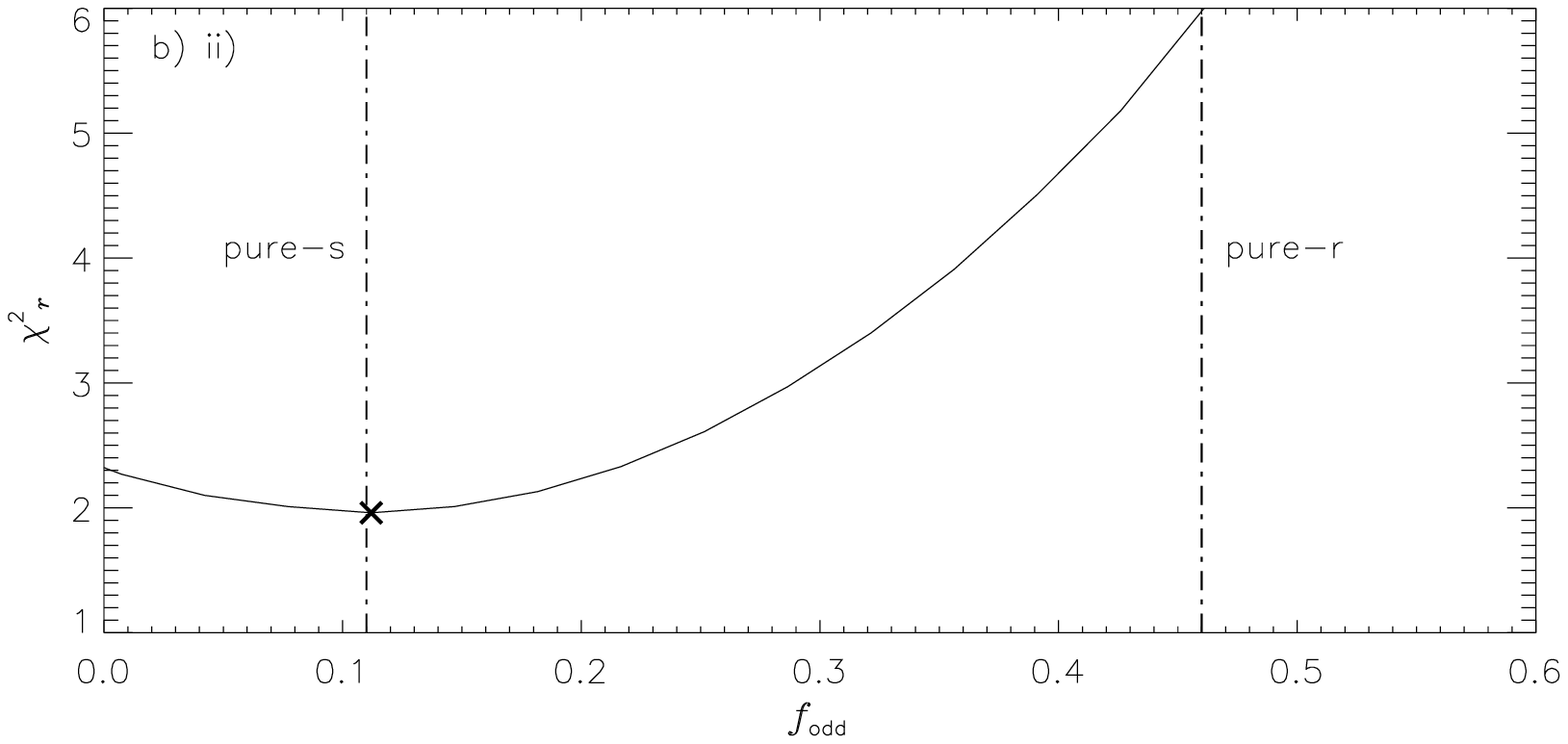}}
	\caption{\textit{Panel a) i)}: The best statistical fit synthetic profile (\textit{solid line}) for the observed \ion{Ba}{ii} 4554\,\AA\ line (\textit{diamonds}) with the residual and $\chi^2$ plots below. For comparison, a pure \textit{r}-process line and residual has been plotted (\textit{dash-dot line}). The value for $A(\element{Ba})$ has been optimised to one that minimises $\chi^2$, values for $A(\element{Fe})$ and macroturbulence remain the same. \textit{Panel a) ii)}: The $\chi^2$ fit for the 4554\,\AA\ line, the cross shows where the minimum of the fit lies. Also plotted are the splitting patterns for barium relative to barium-138 (see Table \ref{tab:linelist}). \textit{Panels b) i)} and \textit{ii)}: Show the same as \textit{a) i)} and \textit{ii)} for the 4934\,\AA\ line.}
	\label{fig:baspec}
	\end{center}
\end{figure*}

The reduced chi-squared values, $\chi_r^2$, for the 4554\,\AA\ line and the 4934\,\AA\ line are 6.6 and 2.0 respectively. The two best statistical fits for both lines and their residuals (observed - synthetic profiles) are shown in Fig. \ref{fig:baspec}. We also plot the synthetic profiles for the \element{Ba} lines with an \textit{r}-process contribution of 100\%, best fit $A(\element{Ba})=-1.30$ for both lines. It can be seen in the residual plots for both lines that the pure \textit{r}-process fits are very poor. In 4554\,\AA, $\chi^2_r$ changes faster with $\fodd$ than for the 4934\,\AA\ line. This indicates that although the 4934\,\AA\ line is broader due to the effects of hyperfine structure, the 4554\,\AA\ line is more sensitive to changes in $\fodd$. This could be both because the 4554\,\AA\ line is stronger ($W = 20.1$\,m\AA) than the 4934\,\AA\ line ($W = 13.6$\,m\AA), and because the latter has an \element{Fe} blend. 

\begin{figure}[!ht]
	\begin{center}
	\resizebox{\hsize}{!}{\includegraphics{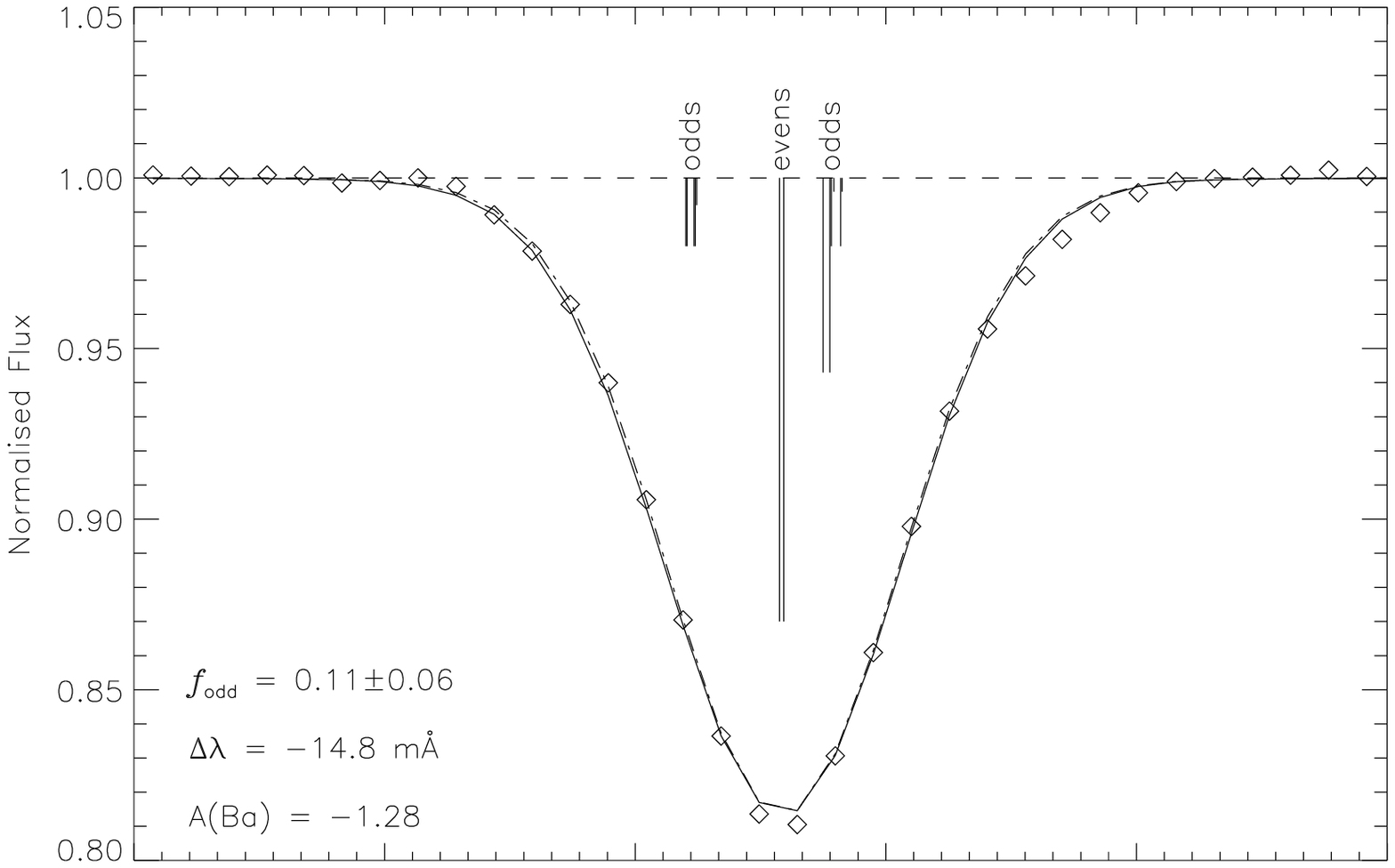}}
	\resizebox{\hsize}{!}{\includegraphics{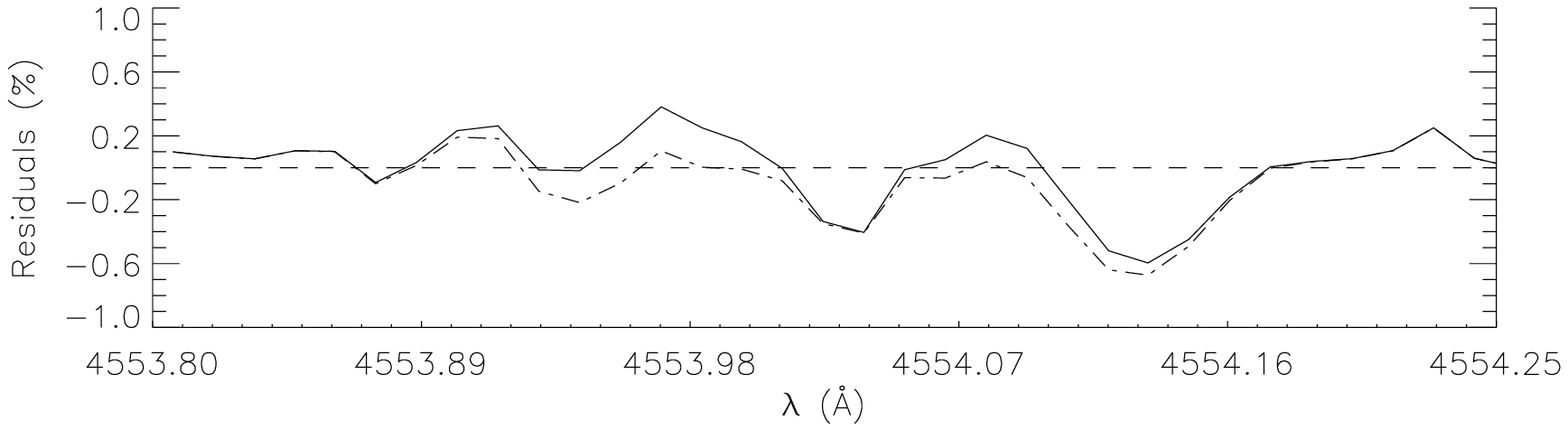}}
	\caption{Comparison between the nearest physical fit (\textit{solid line}) where $\fodd=0.11$ and the best statistical fit (\textit{dash-dot line}) where $\fodd=0.01$, for the 4554\,\AA\ line. Also plotted are the splitting patterns for barium relative to barium-138 (see Table \ref{tab:linelist}).}
	\label{fig:4554real}
	\end{center}
\end{figure}

Based on the calculations by \citet{Arlandini99}, our 4554\,\AA\ result should not be achievable, and corresponds to an \textit{r}-process contribution of -29\% (i.e. the \textit{s}-process contribution is equal to 129\%). We have also plotted in Fig. \ref{fig:4554real} the fit and residual for the nearest physically possible value for $\fodd$ (0.11). This fit has $\chi^2_r = 7.7$. We have also plotted the fit for $\fodd = 0.01$, which is quite similar.


\begin{table}[!ht]
\begin{center}
\begin{threeparttable}
\caption{Results from previous studies of HD\,140283.}
\begin{tabular}{c c c c c c}
\hline\hline	
\vspace{-3mm} \\
Paper					&	$\Teff$ (K)	&	$\logg$ (cgs)	&	[Fe/H]	&	[Ba/H] & [Ba/Fe]				\\
\vspace{-3mm} \\
\hline
\vspace{-3mm} \\
SS\tnote{$a$}		&		5727						&		3.30								&		-2.40	&	-3.20 &	-0.80				\\
MMZ\tnote{$b$} 	&		5640						&		3.10								&		-2.73	&	-3.86 &	-1.13				\\		
GS\tnote{$c$}		&		5690						&		3.58								&		-2.53	&	-3.17 &	-0.64				\\		
RNB\tnote{$d$}	&		5750						&		3.40								&		-2.54	&	-3.45 &	-0.91				\\		
MGB\tnote{$e$}	&		5640						&		3.65								&		-2.30	&	-3.10 &	-0.80				\\		
F\tnote{$f$}		&		5650						&		3.40								&		-2.40	&	-3.43 &	-1.03				\\		
MK\tnote{$g$}		&		5650						&		3.50								&		-2.50	&	-3.28 &	-0.78				\\		
LAP\tnote{$h$}	&		5777						&		3.74								&		-2.70	&	-3.79 &	-1.09				\\		
CAN\tnote{$i$}	&		5690						&		3.67								&		-2.50	&	...   &	...					\\		
GRPA\tnote{$j$}	&		5750						&		3.70								&		-2.59	&	-3.46 &	-0.87				\\		
\hline
\end{tabular}
\begin{tablenotes}
\item[$a$]  \citet{SS78}.
\item[$b$]	\citet{Magain89} and \citet{MZ90}.
\item[$c$]	\citet{GS94}.
\item[$d$]	\citet{RNB96}.
\item[$e$]	\citet{MGB99}.
\item[$f$]	\citet{F00}.
\item[$g$]	\citet{MK01}.
\item[$h$]	\citet{Lambert02}.
\item[$i$]	\citet{Remo09}.
\item[$j$]	This work.
\end{tablenotes}
\label{tab:result}
\end{threeparttable}
\end{center}
\end{table}						

\begin{figure*}[!ht]
	\begin{center}
	\resizebox{0.49\hsize}{!}{\includegraphics{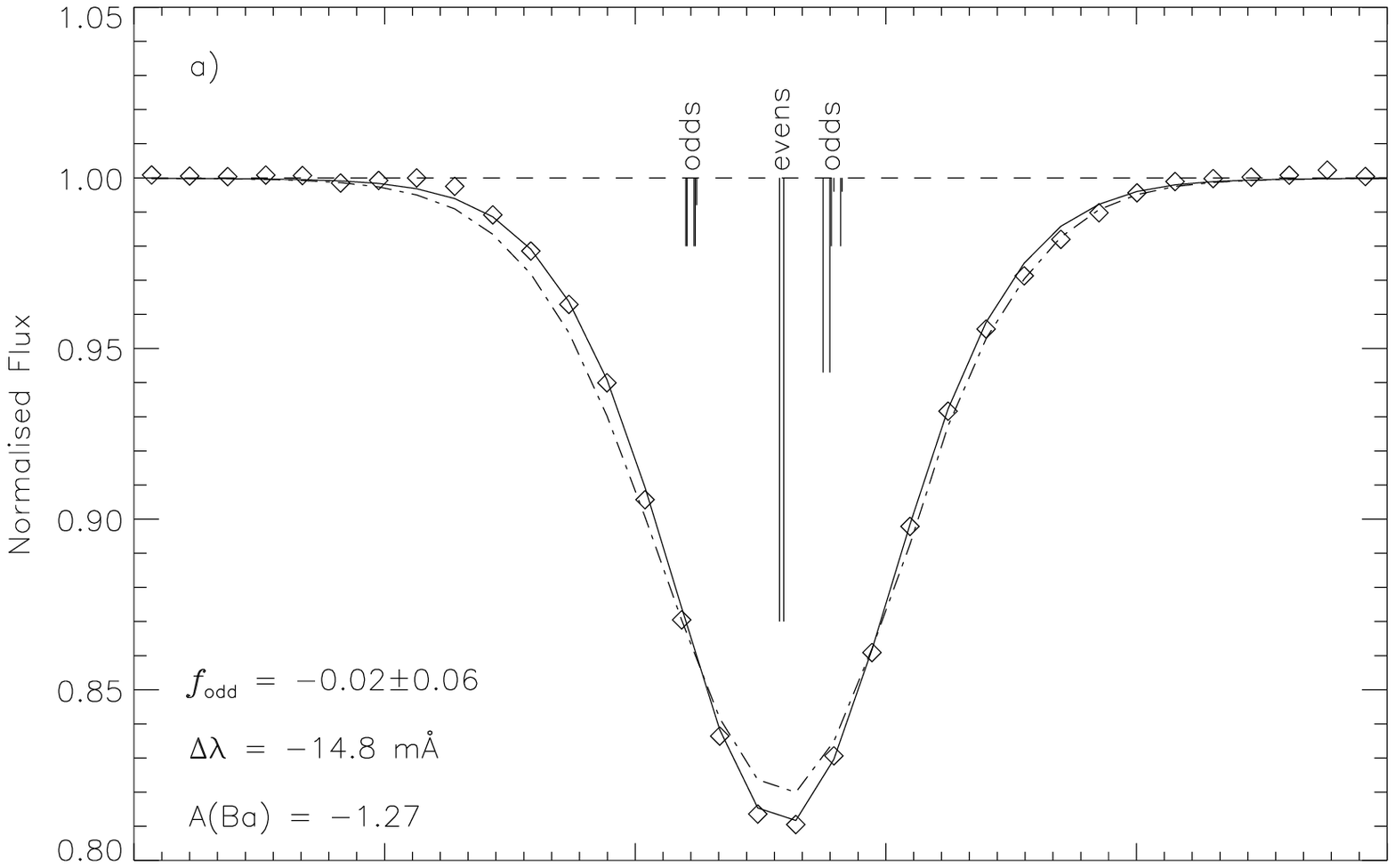}}
	\resizebox{0.49\hsize}{!}{\includegraphics{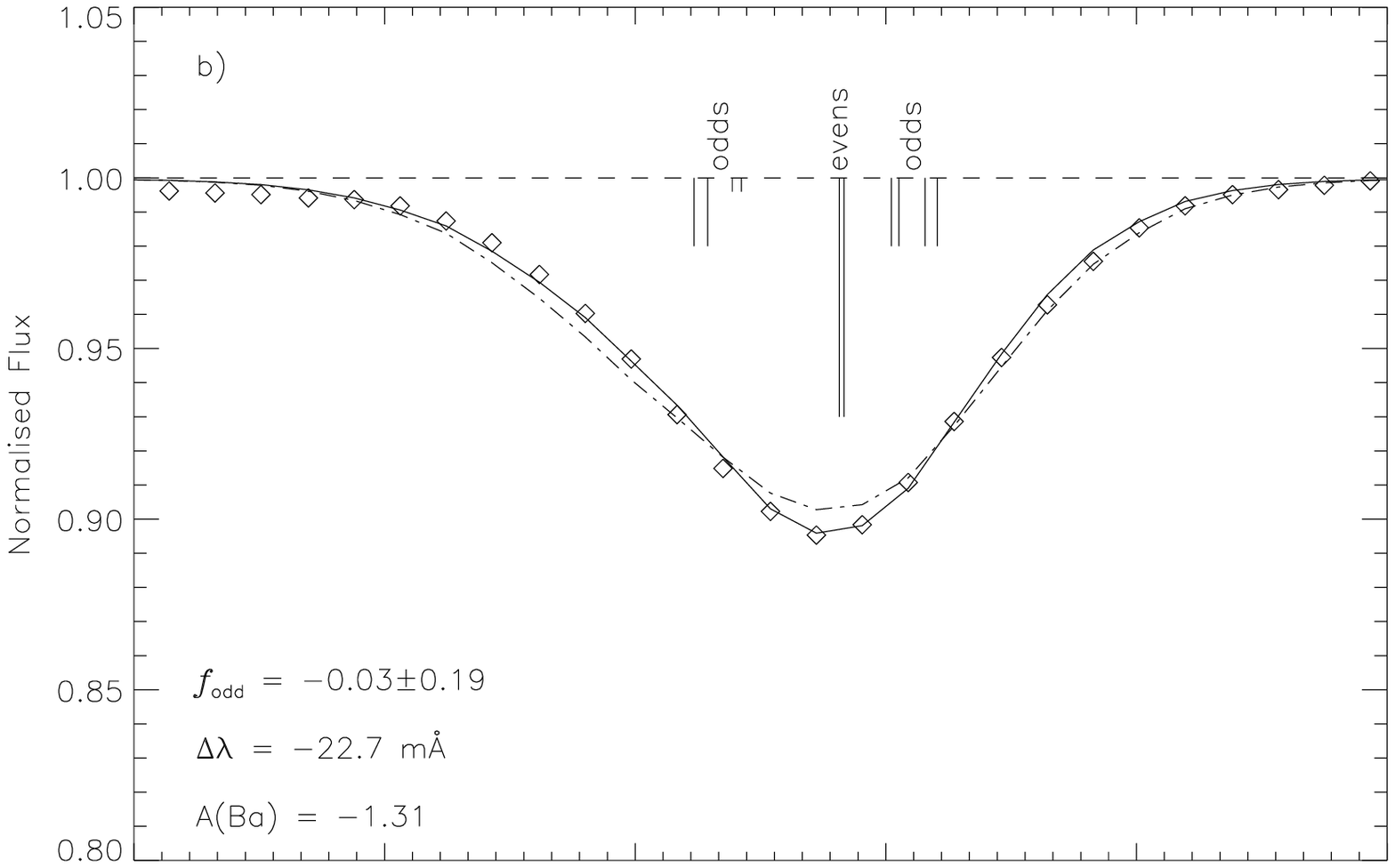}}
	\resizebox{0.49\hsize}{!}{\includegraphics{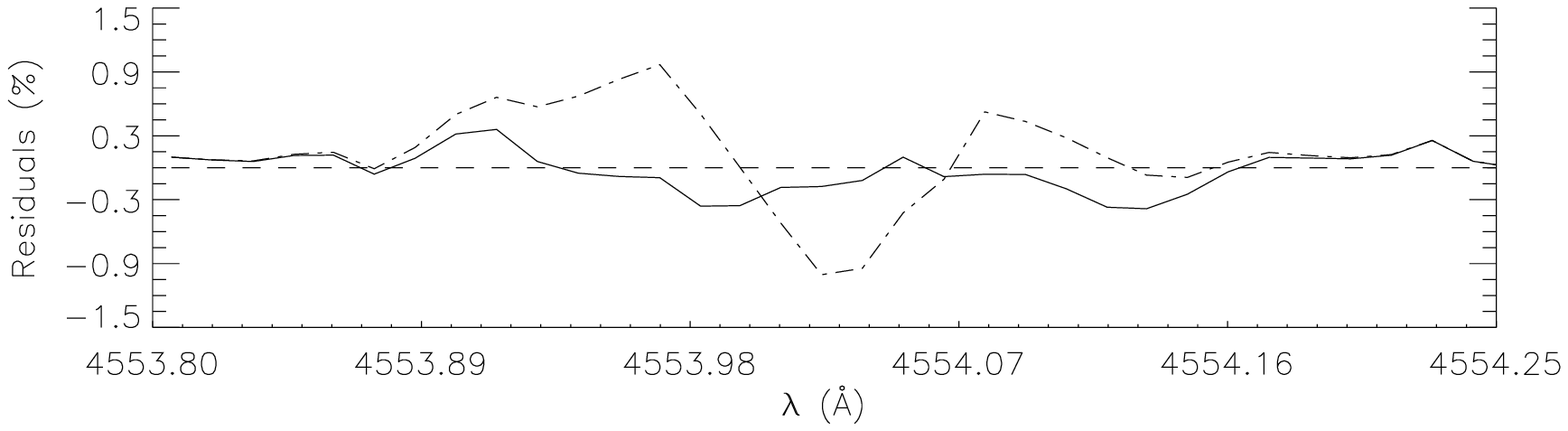}}
	\resizebox{0.49\hsize}{!}{\includegraphics{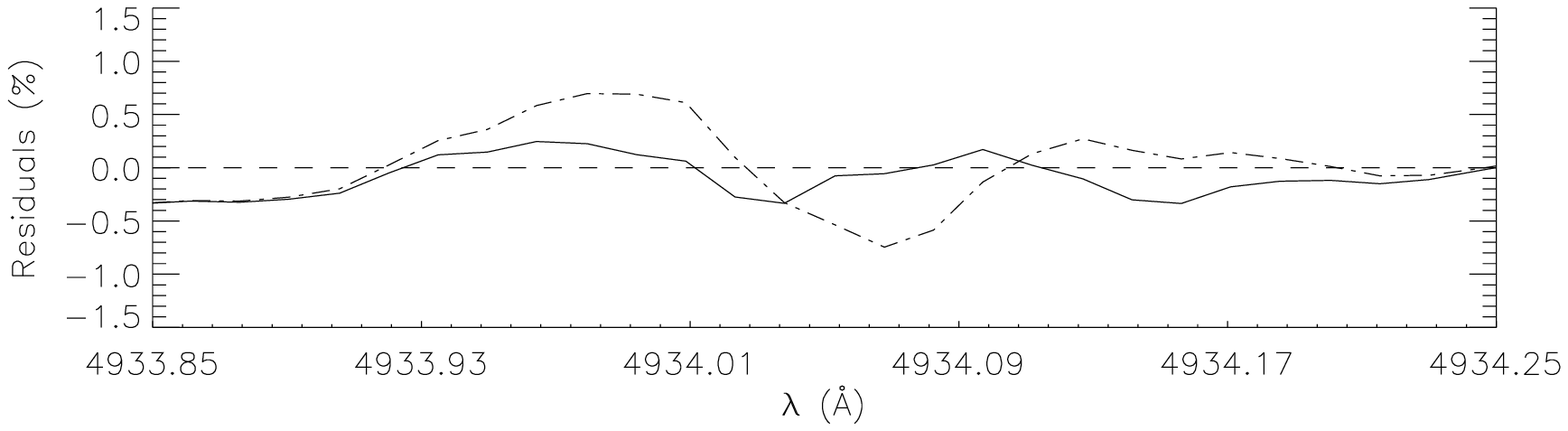}}
	\caption{\textit{Panel a)}: The best statistical fit for the 4554\,\AA\ line (\textit{diamonds}) using a radial-tangential velocity profile (\textit{solid line}). We have included a pure \textit{r}-process, $\fodd = 0.46$, synthetic profile for comparison(\textit{dash-dot line}). Also plotted are the splitting patterns for barium relative to barium-138 (see Table \ref{tab:linelist}). \textit{Panel b)}: Same as \textit{panel a)} but for the 4934\,\AA\ line.}
	\label{fig:baspecrt}
	\end{center}
\end{figure*}

When we adopted radial-tangential macroturbulence, it was determined that $\fodd= -0.02\pm0.06$ and $-0.03\pm0.19$ for the 4554\,\AA\ and 4934\,\AA\ lines respectively, with $\chi^2_r=6.1$ and 2.8. Errors stated here are assumed to be the same as those calculated using a Gaussian macroturbulence as only the broadening mechanism differs; the errors in the two broadening techniques have the same value. The best fits are shown in Fig. \ref{fig:baspecrt}. So the radial-tangential fit for the 4554\,\AA\ line is a statistically better fit than the Gaussian macroturbulent fit, as seen by the residual plots (Figs. \ref{fig:baspec} (a)(i) \& (b)(i) vs. Figs. \ref{fig:baspecrt} (a) \& (b)). Both broadening mechanisms, which were analysed separately, yield similar values for $\Delta\lambda$, $\fodd$ and [Ba/H]. Both indicate a strong \textit{s}-process signature for barium. Although these $\fodd$ numbers are again beyond possible physical values, we inform the reader that due to the finite confidence in the $\chi^2_r$ test (discussed further in \S \ref{sec:discussion}), the unphysical values $\fodd \simeq 0.01$ are not greatly preferred over the physical value $\fodd = 0.11$.

We have given values for $\fodd$ for the two \element{Ba} lines and we have shown that the two lines are in agreement within the stated errors. We now discuss those uncertainties and what stellar parameters $\fodd$ is sensitive to.  

\section{Uncertainties and sensitivity tests}
\label{sec:uncertainty}

\begin{table*}[!ht]																											
\begin{center}																											
\caption{The sensitivity to [Fe/H], [Ba/H], $\nu_{\rm conv}$ and $\fodd$ for different values of $\Teff$ and $\logg$. \textit{case 1}: The sensitivity of all atmospheric parameters used or calculated in this analysis to changes in temperature and $\logg$. \textit{case 2}: The sensitivity of [Ba/Fe] and $\fodd$ to temperature and $\logg$. Macroturbulence and [Fe/H] are fixed to values found for the model atmosphere used in this analysis ($\Teff = 5750$, $\logg = 3.7$, $\nu_{\rm conv} = 5.75 \kms$, ${\rm [Fe/H]} = -2.59$). Columns (11) and (12) show the sensitivity of the blended 4934\,\AA\ line to changes in the stellar parameters.}
\begin{tabular}{c	c	c	c	c	c c c c c c c c c c c}																						
\hline\hline
\vspace{-3.25mm} \\								
																											
				&					&					&					&			& 			&  			&	&	        &        &   &   \multicolumn{5}{c}{$\fodd$}	\\
\vspace{-3.25mm} \\								
\cline{12-16}
\vspace{-3mm} \\								
$\Teff$\,(K)	&	$\logg$ (cgs)	&	\multicolumn{2}{c}{$\nu_{\rm conv}$ ($\kms$)}	& &	\multicolumn{2}{c}{[Fe/H]}	& &	\multicolumn{2}{c}{[Ba/H]}	& &	\multicolumn{2}{c}{4554\,\AA}	& &	\multicolumn{2}{c}{4934\,\AA}		\\
\hline
\vspace{-3mm} \\
(1) & (2) & (3) & (4) & & (5) & (6) & & (7) & (8) & & (9) & (10) & & (11) & (12) \\
\hline
				&					&	case 1	&	case 2	&			&	case 1&	case 2&	&	case 1	&	case 2 &  &	case 1& case 2&	& case 1& case 2\\
\cline{3-4} \cline{6-7} \cline{9-10} \cline{12-13} \cline{15-16} \\
\vspace{-6mm}	\\			
5500		&	3.4			&	5.79		&	...			&			&	$-2.78$	&	...	&	&	$-3.69$		&	$-3.71$	 &	&	$-0.01$	&	0.02	&	& 0.15	&	$-0.115$	\\
5500		&	3.7			&	5.67		&	...			&			&	$-2.77$	&	...	&	&	$-3.59$		&	$-3.61$	 &  &	0.05	&	$-0.01$	&	& 0.20	&	$-0.115$	\\
5500		&	4.0			&	5.52		&	...			&			&	$-2.74$	&	...	&	&	$-3.50$		&	$-3.50$	 &	&	0.11	&	$-0.07$	&	& 0.24	&	$-0.111$	\\
\vspace{-2mm} \\
5750		&	3.4			&	5.82		&	...			&			&	$-2.60$	&	...	&	&	$-3.50$		&	$-3.52$	 &	&	$-0.02$	&	0.04	&	& 0.09	&	0.113	\\
5750		&	3.7			&	5.75		&	5.75		&			&	$-2.59$	&	$-2.59$	&	&	$-3.42$		&	$-3.42$	 &	&	0.01	&	0.01	&	& 0.11	&	0.112	\\
5750		&	4.0			&	5.60		&	...			&			&	$-2.57$	&	...	&	&	$-3.33$		&	$-3.31$	 &	&	0.09	&	$-0.03$	&	& 0.15	&	0.112	\\
\vspace{-2mm} \\
6000		&	3.4			&	5.84		&	...			&			&	$-2.43$	&	...	&	&	$-3.34$		&	$-3.34$	 &	&	$-0.03$	&	0.04	&	& 0.02	&	0.252	\\
6000		&	3.7			&	5.77		&	...			&			&	$-2.42$	&	...	&	&	$-3.24$		&	$-3.24$	 &	&	0.01	&	0.02	&	& 0.04	&	0.252	\\
6000		&	4.0			&	5.69		&	...			&			&	$-2.40$	&	...	&	&	$-3.15$		&	$-3.14$	 &	&	0.04	&	0.00	&	& 0.05	&	0.245	\\
\hline																											
\end{tabular}
\label{tab:temp}																											
\end{center}																											
\end{table*}

In this section we scrutinise the analysis procedures and statistical tests employed in this study to determine the likely statistical and systematic errors. These include errors associated with the atmospheric parameters used in constructing the synthetic spectra, the calculated macroturbulence and the errors associated with the iron lines used in conjunction with the \element{Ba} 4934\,\AA\ line. 

Table \ref{tab:temp} lists values found for [Ba/H] and $\fodd$ by varying the temperature and $\logg$ of the model atmosphere. There are two cases. In case 1 we recalculate [Fe/H] and macroturbulence for every value of $\Teff$ and $\logg$, whereas in case 2 we fix the macroturbulence and [Fe/H] to values calculated for $\Teff = 5750$\,K and $\logg = 3.7$. Perhaps the first thing to note from this table is that altering temperature by $\pm250$\,K and $\logg$ by $\pm0.3$ does not drive $\fodd$ to an \textit{r}-process dominated fraction. The errors quoted in this paper for HD\,140283 are based on uncertainties in $\Teff$ and $\logg$ of $\pm100$\,K and $\pm0.1$ respectively.

From Table \ref{tab:temp} it is possible to calculate the error associated with [Fe/H], $A(\element{Ba})$, and hence [Ba/H] \& [Ba/Fe], by examining how it is affected by the stellar parameters. We use case 1 to calculate these uncertainties. It is shown in Table \ref{tab:temp} that gravity as very little affect on [Fe/H]. Temperature has a much greater affect on [Fe/H], altering the ratio by $\pm0.07$\,dex for every 100\,K. Therefore we find a total uncertainty in [Fe/H] of $\pm0.07$. We find that $\delta A(\element{Ba})/\delta\logg = 0.35$. Therefore an error of 0.1 in $\logg$ implies an error $\sigma_{A(\element{Ba}),\logg}=0.04$. Similarly we find for temperature that $\delta A(\element{Ba})/\delta\Teff=0.0007 \ {\rm K}^{-1}$. Taking the uncertainty in temperature to be $\pm 100$\,K we find that $\sigma_{A(\element{Ba}),\Teff}=0.07$. Macroturbulence affects the shape of lines but not the equivalent width. As such we do not include the uncertainties associated with macroturbulence here. Also [Fe/H] has very little affect on [Ba/H] when compared to temperature and gravity effects so this is not included in our error analysis of [Ba/H]. The solar barium abundance is $A(\element{Ba})_{\sun}=2.17\pm0.07$ \citep{Grevesse98}. When these uncertainties are added in quadrature we find that ${\rm [Ba/H]}=-3.46\pm 0.11$. Therefore we find that ${\rm [Ba/Fe]}=-0.87\pm 0.14$.

\subsection{The 4554\,\AA\ line}
\label{sec:4554uncert}

We have stated that potentially the most significant parameter that $\fodd$ is sensitive to is macroturbulence. The difference between columns 9 and 10 in Table \ref{tab:temp} show how $\fodd$ is sensitive to macroturbulence. We find that on average, $\delta\fodd/\delta\nu_{\rm conv}\backsimeq -0.7 \ ({\rm km \ s^{-1}})^{-1}$ meaning that for $\sigma_{\nu_{\rm conv}} = 0.02\kms$, calculated in \S \ref{sec:macroturbulence}, $\sigma_{\fodd}\backsimeq0.01$. That is, by using a large number of \element{Fe} lines to constrain $\nu_{\rm conv}$, we have minimized the impact of this error.

Realistically, when you vary one parameter you alter all other parameters to compensate for this change. Increasing temperature increases the derived macroturbulence, which on its own decreases $\fodd$. For case 1 we see that an uncertainty in temperature of $\pm100$\,K implies an uncertainty in macroturbulence of $0.01\kms$ to $0.02\kms$ with increasing gravity (see Table \ref{tab:temp}, column (3)). Using the relation we found between $\fodd$ and macroturbulence we see that $\sigma_{\fodd}=0.02$ to 0.06. In case 2 (where we only look at how $\fodd$ is affected by one stellar parameter) we find that for an error in temperature of $\pm100$\,K, $\sigma_{\fodd}\approx\pm 0.004$ to $0.01$ depending on $\logg$ (Table \ref{tab:temp} column (10), case 2).

The uncertainty in $\logg$ for HD\,140283 is quite small, $\lesssim0.1$, due to its reliable Hipparcos parallax. As gravity affects line broadening, we find that $\logg$ influences the macroturbulence and $\fodd$. Firstly we calculate the effect of $\logg$ on macroturbulence (case 1). We find $\delta\nu_{\rm conv}/\delta\logg\backsimeq 0.4\kms$ depending on temperature. So for an uncertainty in $\logg = 0.1$, $\sigma_{\nu_{\rm conv}}\backsimeq 0.04\kms$. Using the sensitivity we calculated for macroturbulence suggests an uncertainty in $\fodd\backsimeq 0.03$. In comparison, the total case 1 sensitivity is $\delta\fodd/\delta\logg\approx0.2$ implying $\sigma_{\fodd}=0.02$. We can see from Table \ref{tab:temp} the separate effect that $\logg$ has on $\fodd$ when we fix macroturbulence (case 2). Here we find that $\delta\fodd/\delta\logg\backsimeq -0.17$ meaning that an error in $\logg$ of 0.1 alters $\fodd$ directly by 0.02. The implication is that some of the change in case 1 is driven by the revision of the macroturbulence, and some is driven more directly but in a way that partially compensates.

When examining the effect of microturbulence on $\fodd$ one would expect to see two things. If we allow macroturbulence to compensate for the change in microturbulence (case 1) we would expect find that $\fodd$ is essentially unchanged. If we fix macroturbulence and alter microturbulence, $\fodd$ will change. Table \ref{tab:mic} shows these two cases. As expected in case 1, the macroturbulence is driven up/down when the microturbulence is decreased/increased and $\fodd$ is unaffected. In case 2 we see the sensitivity in $\fodd$ as microturbulence is altered given by $\delta\fodd/\delta\xi = -0.5 \ (\rm km \ s^{-1})^{-1}$. Therefore an uncertainty in microturbulence of $0.1\kms$ implies an error in $\fodd = 0.05$.  It is case 1 that is relevant to our \element{Ba} analysis. 

\begin{table}[!ht]																											
\begin{center}																											
\caption{The sensitivity of $\fodd$ to $\xi$. Temperature and $\logg$ are fixed at $5750 \ K$ and 3.7 respectively. \textit{case 1}: The sensitivity of $\fodd$ to microturbulence when macroturbulence is re-evaluated to compensate for the change to microturbulence. \textit{case 2}: The sensitivity of $\fodd$ to microturbulence when macroturbulence is fixed at the value calculated when $\xi = 1.4$.}
\begin{tabular}{@{}c c c c c c c c c@{}}																						
\hline\hline
\vspace{-3.25mm} \\																																	
	$\xi$		&	\multicolumn{2}{c}{$\nu_{\rm conv}$	}        &    &	\multicolumn{5}{c}{$\fodd$}								\\
\vspace{-3.25mm} \\								
\cline{5-9}								
\vspace{-3.25mm} \\								
$(\kms)$ &	\multicolumn{2}{c}{($\kms$)}	& &	\multicolumn{2}{c}{4554\,\AA}	&&	\multicolumn{2}{c}{4934\,\AA}	\\
\hline
			& case 1 & case 2 &     & case 1  & case 2 && case 1 & case 2  \\
         \cline{2-3}              \cline{5-6}           \cline{8-9} \\
\vspace{-6mm}	\\			
1.3		&	5.81	 & ...	  &     & 0.01	 & 0.06   &&	0.11   & 0.14		\\
1.4		&	5.75	 & 5.75		&     &	0.01	 & 0.01   &&	0.11   & 0.11		\\
1.5		&	5.68	 & ...	  &     &	0.01	 & -0.04  &&	0.11   & 0.07		\\
\hline																											
\end{tabular}
\label{tab:mic}																											
\end{center}																											
\end{table}

In summary we can assign an uncertainty in $\fodd$ for the 4554\,\AA\ line $\sigma^2_{\fodd}=\sigma^2_{\nu_{\rm conv}}+\sigma^2_{\Teff}+\sigma^2_{\logg}=\sqrt{0.01^2+0.02^2+0.03^2} = 0.04$ (case 1 - remember that $\nu_{\rm conv}$ compensates for any effect $\xi$ has on $\fodd$). In case 2, where we look at the separate effects the stellar parameters have on $\fodd$, we find that for uncertainties in macroturbulence, temperature, $\logg$ and microturbulence $\sigma_{\fodd}=\sqrt{0.01^2+0.01^2+0.02^2+0.05^2} = 0.06$. Case 1 is probably more applicable, but we adopt the larger error, case 2, as a precaution, i.e. $\pm0.06$. We now move on to errors and uncertainties associated with the 4934\,\AA\ line.

\subsection{The 4934\,\AA\ line}
\label{sec:4934uncert}

In order to assign an uncertainty in $\fodd$ to the 4934\,\AA\ line we must also explore how uncertainties in the \element{Fe} blend (see Table \ref{tab:feblend}) affect $\fodd$.

We explored how the equivalent width of the \element{Fe} blend is affected by temperature and $\logg$. As in Tables \ref{tab:temp} and \ref{tab:mic}, we computed two cases where we allow macroturbulence and [Fe/H] to vary with varying temperature and $\logg$ (case 1) and where we have fixed macroturbulence and [Fe/H] (case 2) - in Table \ref{tab:temp}. 

The net result of an increase in $\logg$, decrease in macroturbulence, and increase in [Fe/H], is a small increase in synthesized $W_{\rm Fe}$, but these effects are minimal compared to the effects that macroturbulence has on $\fodd$. Consequently we see a similar behaviour in $\fodd$ (Table \ref{tab:temp}, column (11)) as that exhibited by the 4554\,\AA\ line (column (9)) (a roughly linear increase in $\fodd$ with $\logg$ with $\Delta{\fodd}_{,4934}$ comparable to $\Delta{\fodd}_{,4554}$).

In case 2 an increasing temperature decreases the equivalent widths of the \element{Fe} lines. Unlike case 1, macroturbulence and $A(\element{Fe})$ are not compensating for the increasing ionisation fraction meaning that \ion{Fe}{i} level populations are decreasing. This decreases the strength of the \element{Fe} lines, decreasing their equivalent widths. 

As $\logg$ is driven up in case 2, we find that the equivalent widths are decreasing, recall that $A(\element{Fe})$ is fixed in case 2. Overall, however, we see little or no change in $\fodd$ in column (12) in Table \ref{tab:temp}.

\begin{table*}[!t]
\begin{center}
\caption{Values of $\fodd$ for the 4934\,\AA\ line found from altering the strength of the \element{Fe}  lines. By changing the $\loggf$ values by $\pm 0.15$ we find that $\fodd$ alters to compensate for a weakening/strengthening of the \element{Ba} wing. We have also included an extra decimal place to show that there is a slight change in $\fodd$ as we alter the temperature and gravity in case 2.}
\begin{tabular}{@{}ccccccccccccccccccc@{}}
\hline\hline
$\Teff$\,(K)&$\logg$ (cgs)&\multicolumn{2}{c}{$\nu_{\rm conv} \ (\kms)$}&&\multicolumn{2}{c}{[Fe/H]}&& \multicolumn{5}{c}{[Ba/H]}&&\multicolumn{5}{c}{ $\fodd$ (4934\,\AA)}\\
\hline
\vspace{-3mm} \\
(1) & (2) & (3) & (4) & & (5) & (6) & & (7) & (8) & & (9) & (10) & & (11) & (12) & & (13) & (14) \\
\hline
     &       &	case 1 & case2 && case1 & case2 &&\multicolumn{2}{c}{case 1}&&\multicolumn{2}{c}{case 2}&&\multicolumn{2}{c}{case 1}&&\multicolumn{2}{c}{case 2}\\
\vspace{-4mm} \\
\cline{3-4} \cline{6-7} \cline{9-13} \cline{15-19} \\
\vspace{-6mm} \\
     &	     & 	       &       &&       &       &&-0.15&+0.15&&-0.15&+0.15&&-0.15&+0.15&&-0.15&+0.15\\
\cline{9-10} \cline{12-13} \cline{15-16} \cline{18-19} \\
\vspace{-6mm} \\
5500 &	3.4  &	5.79   & ...  && $-2.78$ & ... && $-3.67$ & $-3.73$ && $-3.71$ & $-3.81$ && 0.302  &  $-0.046$ && 0.126  &  ...    \\
5500 &	3.7  &	5.67   & ...  && $-2.77$ & ... && $-3.58$ & $-3.63$ && $-3.60$ & $-3.69$ && 0.342  &  $-0.007$ && 0.127  &  ...    \\
5500 &	4.0  &	5.52   & ...  && $-2.74$ & ... && $-3.48$ & $-3.53$ && $-3.49$ & $-3.57$ && 0.376  &   0.043 && 0.126  &  ...    \\
\vspace{-2mm} \\
5750 &	3.4  &	5.82   & ...  && $-2.60$ & ... && $-3.51$ & $-3.56$ && $-3.51$ & $-3.56$ && 0.252  &  $-0.150$ && 0.275  & $-0.134$  \\
5750 &	3.7  &	5.75   & 5.75 && $-2.59$ & $-2.59$ && $-3.41$ & $-3.47$ && $-3.41$ & $-3.47$ && 0.273  &  $-0.127$ && 0.273  & $-0.127$  \\
5750 &	4.0  &	5.60   & ...  && $-2.57$ & ... && $-3.32$ & $-3.37$ && $-3.31$ & $-3.36$ && 0.316  &  $-0.100$ && 0.265  & $-0.115$  \\
\vspace{-2mm} \\
6000 &	3.4  &	5.84   & ...  && $-2.43$ & ... && $-3.34$ & $-3.39$ && $-3.32$ & $-3.36$ && 0.200  &  ...&& 0.373  &  0.083  \\
6000 &	3.7  &	5.77   & ...  && $-2.42$ & ... && $-3.25$ & $-3.31$ && $-3.23$ & $-3.27$ && 0.221  &  ...&& 0.369  &  0.084  \\
6000 &	4.0  &	5.69   & ...  && $-2.40$ & ... && $-3.15$ & $-3.21$ && $-3.13$ & $-3.17$ && 0.235  &  ...&& 0.362  &  0.083  \\
\hline
\end{tabular}
\label{tab:fesens}
\end{center}
\end{table*}

We also investigate how the $\loggf$ values, which are not well known for the two \element{Fe} lines, affect $\fodd$. The 4934\,\AA\ line is driven to a pure \textit{r}-process fraction if the \element{Fe} blend is eliminated from the line list, so we would expect that $\fodd$ would be quite sensitive to $\loggf$. Table \ref{tab:feblend} shows the parameters of the two lines. We analyse the case that the $\loggf$ values have an error $\pm0.15$ as a heuristic estimate. Table \ref{tab:fesens} shows how $\fodd$ for the 4934\,\AA\ line is affected by this increase/decrease in \element{Fe} strength. We have tabulated the results for $\fodd$ for all values of temperature and $\logg$ we use in our sensitivity analysis. It can be seen for case 2 at a temperature of $5500 \ {\rm K}$ and where the \element{Fe} blend strengths have been increased, that $\fodd$ becomes so small and so unphysical that our $\chi^2$ program cannot find a minimum solution. Similarly this is seen in case 1 at a temperature of $6000 \ {\rm K}$. It is clear from Table \ref{tab:fesens} that for 4934\,\AA, $\fodd$ is more sensitive to the uncertainty in the strengths of the \element{Fe} lines than the atmospheric parameters. We see in case 1 that as we alter the \element{Fe} $\loggf$ by $\pm 0.15$, $\fodd$ is altered by $\mp 0.18$. This means that when added in quadrature to the error discussed in \S \ref{sec:4554uncert}, we find that for case 1, $\fodd = 0.11\pm 0.18$. For case 2, $\fodd = 0.11\pm 0.19$. We take the error to be the average of the two, so $\fodd=0.11\pm0.19$. 

\subsection{Overall result}
\label{sec:result}

Inverse-variance-weighting the results for 4554\,\AA\ (0.01) and 4934\,\AA\ (0.11) give an overall result $\fodd=0.02\pm0.06$ when macroturbulence is modelled as a Gaussian. When a radial-tangential broadening mechanism is used we find that inverse-variance-weighting gives an overall result $\fodd=-0.02\pm0.06$. 

So far the uncertainties discussed in the section have been limited to errors in $\Teff$, $\logg$, $\nu_{\rm conv}$ and $\loggf$. We have not yet quantified the impact of finite $S/N$ and possible systematic errors associated with a 1D LTE analysis. We recall from \S\ref{sec:gaussian} that systematic errors of order 0.09 may arise from using \element{Fe} lines to estimate, in 1D, the macroturbulent broadening of \element{Ba}. We discuss this further in \S \ref{sec:discussion}. We shall now move on and discuss the europium abundance and the various implications of the \element{Ba} and \element{Eu} results.  

\section{Europium abundance limit}
\label{sec:[Ba/Eu]}

\begin{figure*}[t]
	\begin{center}
  	\resizebox{0.48\hsize}{!}{\includegraphics{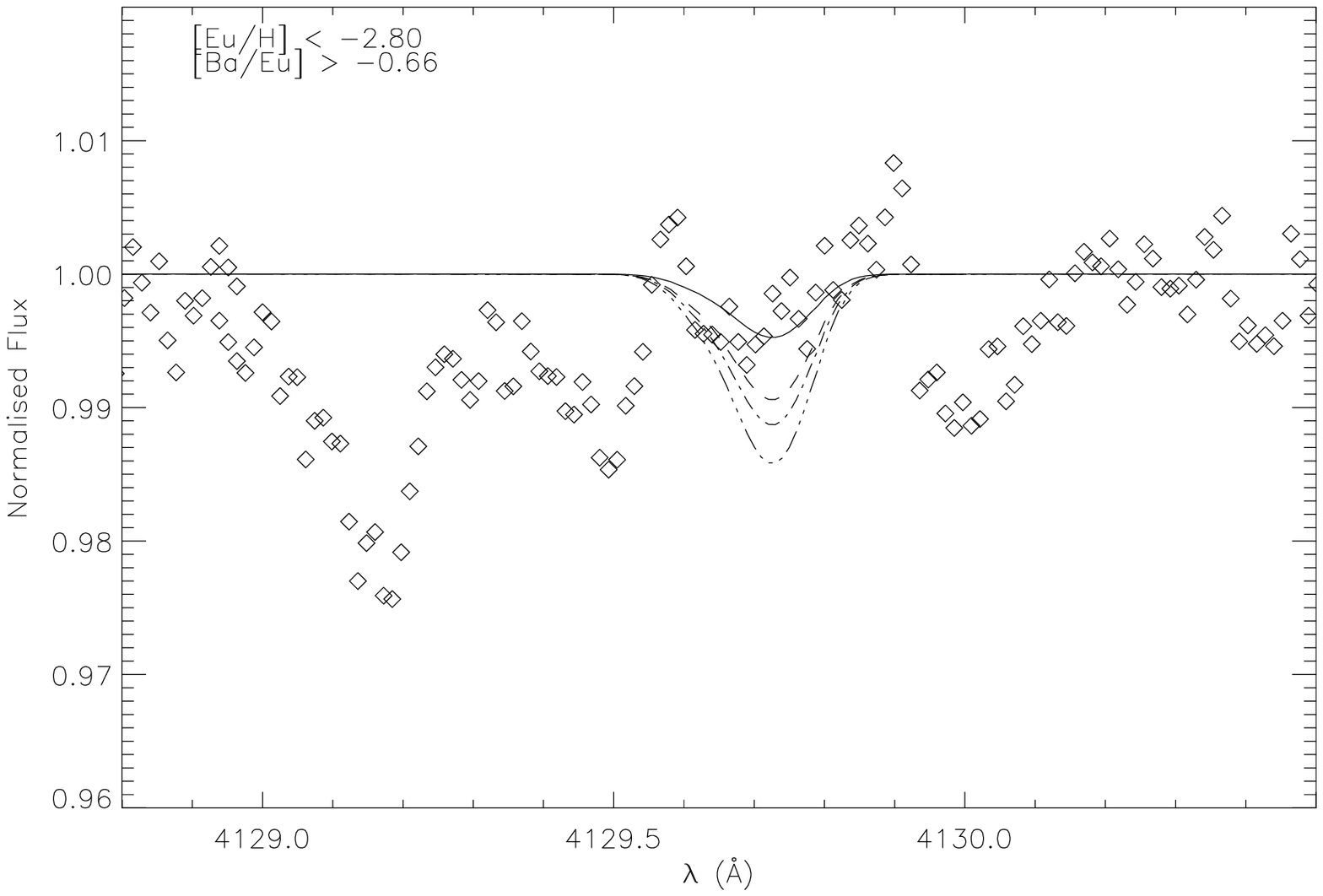}}
		\resizebox{0.48\hsize}{!}{\includegraphics{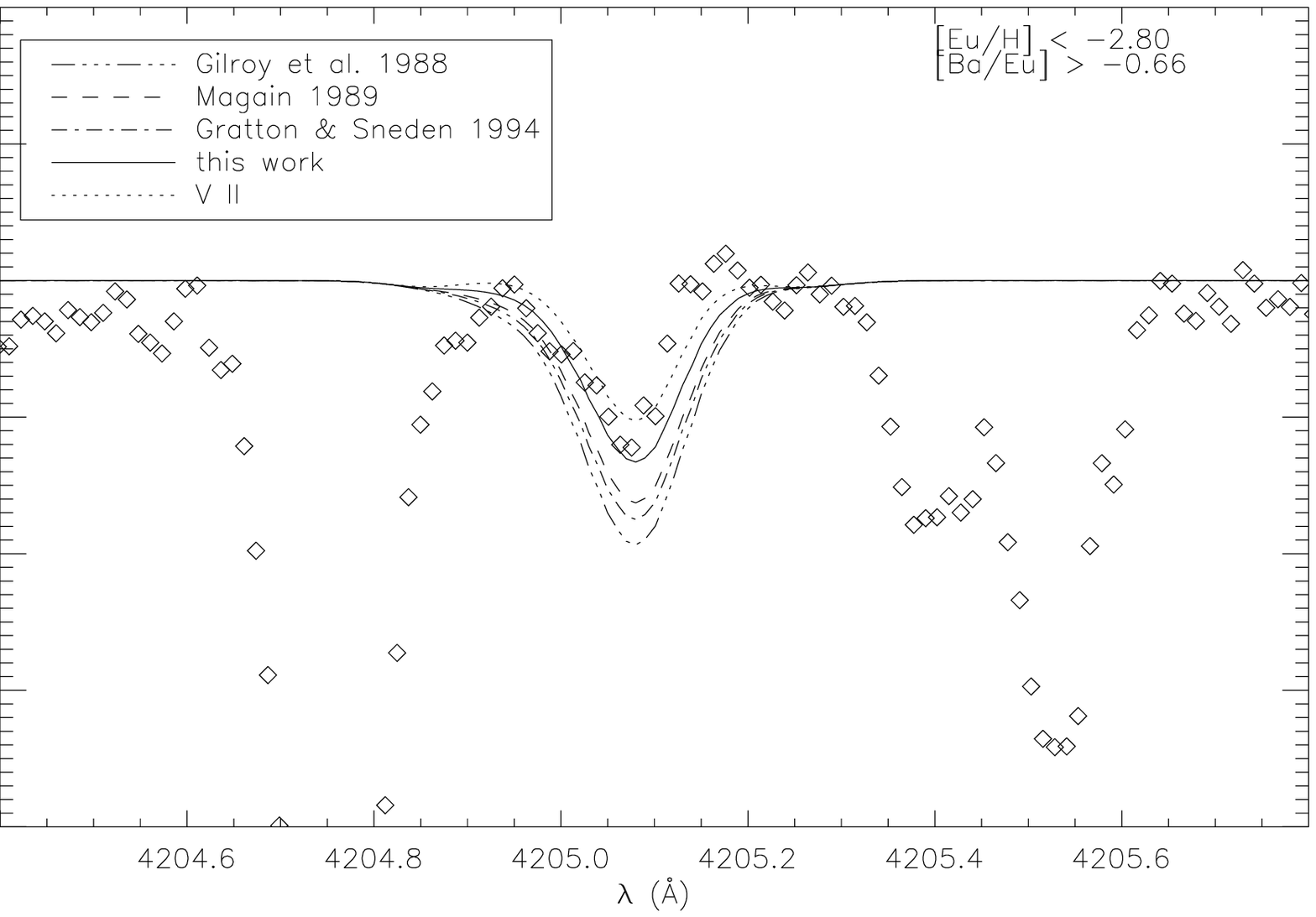}}
	  \caption{Synthetic spectra for the \element{Eu} 4129 and 4205\,\AA\ lines for ${\rm [Eu/H]}=-2.80$ and for the abundances calculated in three other studies of this star (see text). It is clear that they over estimate the strength of both lines. While the 4205\,\AA\ line includes several blends particularly \ion{V}{ii} 4205.09\,\AA, the 4129\,\AA\ has none. We show the \ion{V}{ii} line separately in the right-hand panel (\textit{dot-line}).}
	\label{fig:eulines}
	\end{center}
\end{figure*}

Within our spectral range ($4118-6253$\,\AA) there are two \ion{Eu}{ii} resonance lines, 4129.70\,\AA\ and 4205.05\,\AA. \citet{Honda06} report that the latter has a known blend with a \ion{V}{i} line. \citet{Gilroy88}, \citet{Magain89} and \citet{GS94} report [Eu/H] to be $-2.31$, $-2.49$ and $-2.41$ respectively. However, these lines do not appear strongly in our spectrum. This becomes clear when studying Fig. \ref{fig:eulines}, which presents the observed data and several synthetic spectra. The \element{Eu} line lists were constructed using hyperfine splitting information from \citet{Krebs60} and \citet{Becker93}. We acknowledge that more recent hyperfine splitting information is available from \citet{Lawler01}, which is in good agreement with \citet{Krebs60} and \citet{Becker93} but we do not use that data here. An isotopic ratio of 0.5:0.5 for \element{Eu} 151:153 was chosen for the \textit{r}- and \textit{s}-processes (the solar system isotopic ratio of \element{Eu} 151:153 is 0.48:0.52 \citep{Arlandini99}), and \textit{gf} values from \citet{Biemont82} and \citet{Karner82} were used. The synthetic spectra were produced using {\small KURUCZ06} model atmospheres in conjunction with the 1D LTE code {\small ATLAS}. The macroturbulence and [Fe/H] was set at values calculated in \S \ref{sec:gaussian}.  
 
The \element{Eu} 4205\,\AA\ synthesis includes several blends which we have adopted from the Kurucz theoretical database ({\tiny\texttt{http://www.cfa.harvard.edu/amp/ampdata/kurucz23/sekur.html}}). The dominant blend is a \ion{V}{ii} line at $\lambda=4205.09$\,\AA, which has a \textit{gf} value of 0.089 and $\chi=2.04$\,eV. The abundance of \element{V} was taken as the solar abundance scaled to the metallicity. We found no blends associated with the 4129\,\AA\ line. As these line blends are theoretical, we make no claim that the abundances we deduce from the \element{Eu} analysis are as accurate as the analysis conducted on the \element{Ba} lines. We interpret the absorption feature at 4205.1\,\AA\ as due to \ion{V}{ii}, not \element{Eu}, as it is much narrower than the synthesised, hfs-broadened \element{Eu} line. Moreover if it were \element{Eu}, not \element{V}, it would require an abundance inconsistent with the weakness of the \element{Eu} 4129\,\AA\ line. We find that a [Eu/H] abundance of $-2.80$ seems to be a generous upper limit on the \element{Eu} abundance, rather than a genuine detection and lower than the cited detections. Therefore we assign a lower limit ${\rm [Ba/Eu]} > -0.66$. This marginally excludes an \textit{r}-process ratio, whether we assume an \textit{r}-process limit set in \citet{Burris00} ($-0.81$, which were calculated using the \citet{Andersetal89} isotopic abundances) or \citet{Arlandini99} ($-0.69$). A pure \textit{s}-process ratio ($+1.45$, \citep{Burris00} or $+1.13$, \citep{Arlandini99}) or a mixed \textit{s}- and \textit{r}-process regime, is compatible with the data. However, our [Ba/Eu] limit does agree well with observations found in \citet{Francois07} for stars of similar metallicity to HD\,140283.
     
We shall now move on and discuss the various implications of the results found in this paper and look at possible solutions to reduce systematic errors associated with a 1D LTE analysis. 

\section{Discussion}
\label{sec:discussion}

We have found that $\fodd=0.02\pm0.06$, and hence the \textit{r}-process fraction implies a purely \textit{s}-process signature of \element{Ba} in HD 140283. The [Ba/Eu] ratio, $>-0.66$, is also marginally inconsistent with a pure \textit{r}-process regime. The isotope result does not entirely contradict previous work by \citet{Lambert02} and \citet{Remo09} since, due to the size of their $1\sigma$ errors, an \textit{s}- or \textit{r}-process isotopic mixture was feasible (see Fig. \ref{fig:r/fodd}). Although we find that $\fodd$ for the 4554\,\AA\ line is unphysical (at the $1.8\sigma$ level) based on \citet{Arlandini99}, we must consider the possibility that the adopted \textit{s}- and \textit{r}-process isotope contributions may not be accurate, as they are based on our simplified understanding of nucleosynthesis, which could be flawed. For example, the \citet{Arlandini99} calculations give a solar-system \textit{r}-process isotopic ratio, and we cannot be certain that this applies in the Galactic halo. However, metal-poor stars with \textit{r}-process enhancements do not least have similar neutron-capture abundance patterns to the sun, e.g. CS\,22892-052 \citep{Sneden96}.

We also question whether the $S/N$ ratio is high enough to measure these fractions accurately, and whether a 1D LTE analysis is an adequate tool in investigating isotopic ratios at these high levels of $S/N$, by looking at the confidence limits of the $\chi^2$ minima, which we now discuss. The fact that our best fitting spectra have $\chi^2_r$ values significantly greater than 1 ($\chi^2_r = 6.6$ and 2.0 for the 4554\,\AA\ and 4934\,\AA\ line respectively) indicates that the $\chi^2$ denominator ($\sigma_i$) is not a good description of the deviation of the model spectrum from the data. We interpret the high $\chi_r^2$ values to indicate that systematic errors are present which exceed the random fluctuations in the signal. This influence is confirmed by inspection of Figs. \ref{fig:baspec}, \ref{fig:4554real} \& \ref{fig:baspecrt}, where it can be seen that the residuals do not oscillate randomly from one pixel to the next but rather seem to meander over a cycle of a few pixels. In short, this tells us that the failure of the model profile to match to the data exceeds the error due to noise (mostly photon noise) in the spectrum, and hence $\sigma_i$ as judged from the $S/N$ underestimates the true residual. From Fig. \ref{fig:baspec} it can be seen that the best fit under-fits the core of the lines in order to fit the wings of the lines better. The \textit{r}-process contributes more to the wings of the line, see Fig. \ref{fig:Split} and Table \ref{tab:linelist}. It is interesting to note that we under-fit the red wing of the 4554\,\AA\ line between 4554.11\,\AA\ and 4554.17\,\AA\ (see Figs. \ref{fig:baspec} \& \ref{fig:4554real}). \citet{Lambert02} and \citet{Remo09} in 1D, see the same residual feature at this wavelength interval. When they reanalysed the line in 3D, \citet{Remo09} appeared to remove the feature in the wing. This would suggest that it is a result of convection in a 3D atmosphere rather than a feature induced by inaccuracies when calculating the isotopic shifts. We suspect that the error arises due to the assumptions used in 1D LTE codes that are unable to correctly model physical conditions in a 3D atmosphere.

\begin{figure}[!ht]
	\begin{center}
  	\resizebox{\hsize}{!}{\includegraphics{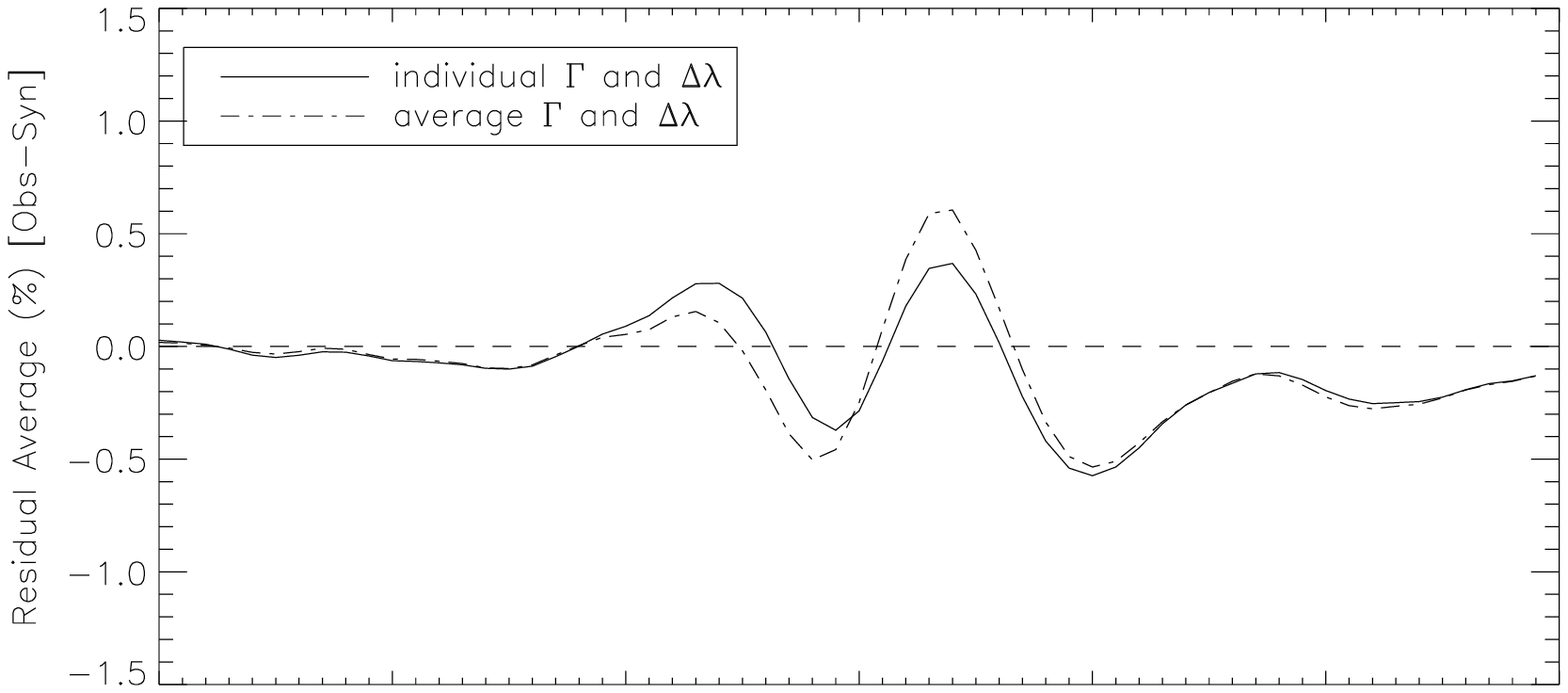}}
		\resizebox{\hsize}{!}{\includegraphics{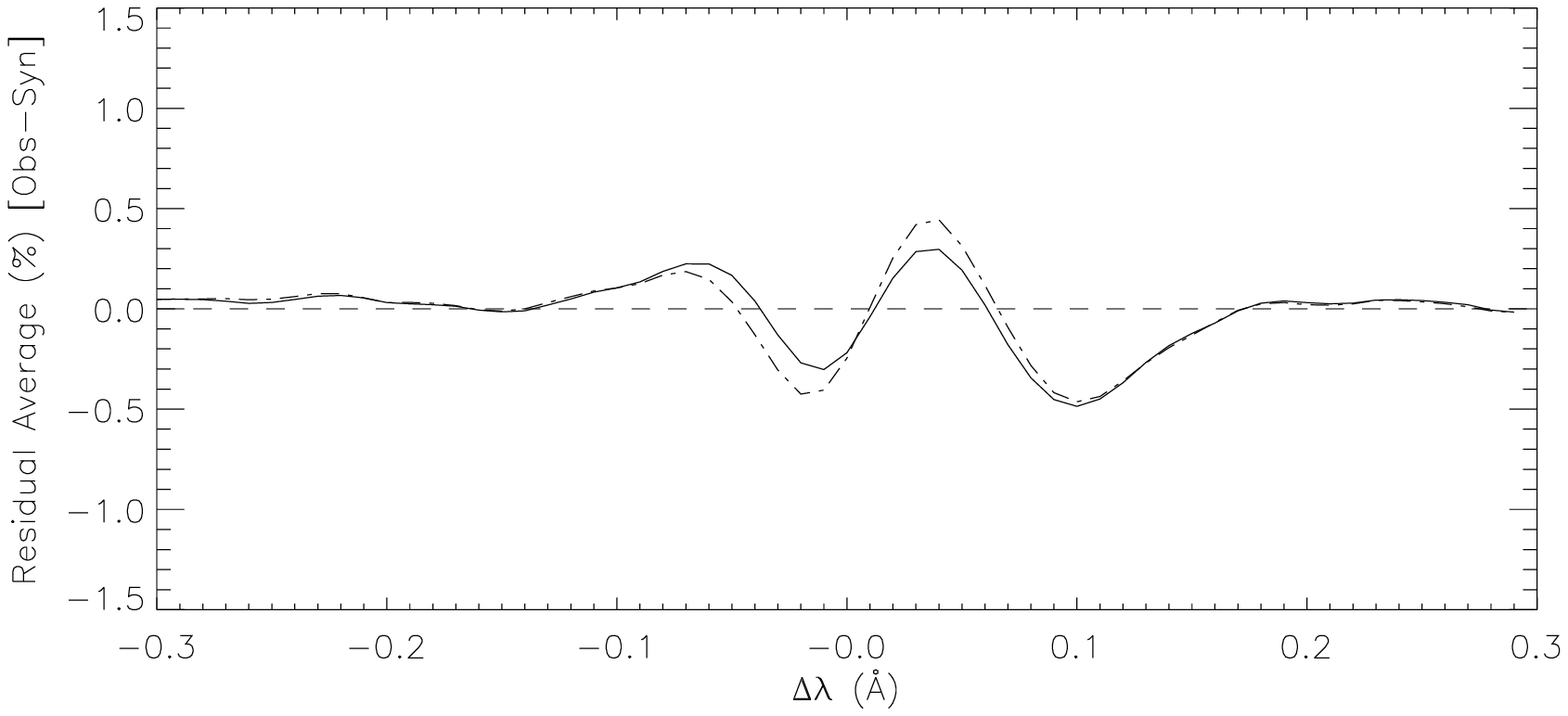}}
	  \caption{Average residuals for fits to 93 \element{Fe} lines, for two cases. It is quite clear that assuming a constant wavelength shift and macroturbulence is adequate when working in 1D LTE but assuming 1D LTE when dealing with high quality data is not. \textit{Top panel}: the reduced-noise residual plot for all 93 lines. \textit{Bottom panel}: the reduced-noise residual plot for 82 lines without other close absorption features. Lines not included in lower panel are noted with an asterisk in Table \ref{tab:Felines}.}
	\label{fig:fecomp}
	\end{center}
\end{figure}

To explore this further, we have searched for evidence of asymmetries in the \element{Fe}-line data. We have produced two \element{Fe}-line plots by co-adding the residuals from all 93 lines to find an average residual, shown in Fig. \ref{fig:fecomp} (\textit{top panel}). The lower panel shows the average residual for the 82 \element{Fe} lines found to have no additional features or close non-\element{Fe} lines within the window over which the $\chi^2$ analysis is calculated (0.6\,\AA). Lines marked with an asterisk in Table \ref{tab:Felines} denote the 11 \element{Fe} lines that were rejected. For one plot (\textit{dash-dot curve}) the average residuals (\textit{obs-syn}) are based on synthetic spectra calculated using the average wavelength shift (-12.04\,m\AA) and average macroturbulence ($5.75\kms$) for all \element{Fe}-lines. The average residual is very asymmetric, in that the blue wings are better fit than the red wings. The large residuals in the red wings, around $60-170$\,m\AA\ from the line centre, are clearly not due to errors in the central wavelengths of the \element{Fe} lines, which are known to better than 1\,m\AA\ (\S \ref{sec:gaussian}).

We also computed a new set of synthetic spectra using the optimised $\nu_{\rm conv}$ and $\Delta\lambda$ values belonging to each \element{Fe} line as found from the $\chi^2$ analysis (see Table \ref{tab:Felines}), and recalculated the residuals (\textit{obs-syn}) for these. These new residuals were coadded and averaged, and plotted in Fig. \ref{fig:fecomp} (\textit{solid curve}). These slightly reduce the overall amplitude of the residuals, as expected since we are optimising the fit to each \element{Fe} line. However, even after allowing each one to be optimised, the residuals are still quite asymmetric, with the red wings standing out as having larger residuals than the blue wings. From Fig. \ref{fig:fecomp} we can see that a similar red feature to the one we see in Fig. \ref{fig:baspec} for \element{Ba} is present in both \element{Fe} residual plots, and appears at the same distance from the centroid of the \element{Fe} lines as the residuals for the \element{Ba} 4554\,\AA\ line ($\sim100$\,m\AA). The feature remains whether we optimise $\nu_{\rm conv}$ and $\Delta\lambda$ or not. We suspect this feature may be the result of convection in the observed, dynamic atmosphere, similar to the feature seen in the \element{Ba} spectrum, that underlying assumptions in 1D LTE do not compensate for. 

\citet{Remo09} conduct a 3D analysis of HD\,140283. Our measurements of the residuals for the \element{Fe} lines provide a future test of whether 3D modelling produces similar residuals. We hope to explore this at a later date. As the 3D process calculates the velocity field ab initio, there is no concept of micro- or macroturbulence in that framework. Consequently, \citet{Remo09} ascribe any excess broadening to stellar rotation, and for this they infer $v\sin{i}=2.5\kms$. We note that the upper limit on $v\sin{i}$ which we infer for zero macroturbulence is $3.9\kms$. Their value is compatible with our limit. We note that \citet{Remo09} find a lower \textit{r}-process fraction for HD\,140283 using a 3D analysis than they find for 1D. If our findings are similar, then this will further accentuate the difference between the analysis of HD\,140283 and the expectations based on Truran's hypothesis. 

One possible alternative explanation of the asymmetries is that we are seeing the combined spectra of more than one star, offset in velocity. It may be difficult to generate the observed levels of asymmetries for a realistic second star, and we have not attempted to do so, but note this possibility nonetheless. We also note that the radial velocity of HD\,140283 has been steady to $\pm0.35\kms$ over long periods of time \citep{Lucatello05}, decreasing the likelihood that it is a binary. 

Finally we shall move on to discuss the conclusions we have drawn from our analysis of the isotopic ratio of \element{Ba} in HD\,140283.

\section{Conclusions}
\label{sec:conclusions}

We have used very high quality data ($S/N=870 - 1110$, $R\equiv\lambda/\Delta\lambda=95 \ 000$) to analyse the \element{Ba} isotopic fraction and the \element{Eu} abundance limit in the metal-poor subgiant HD\,140283. We obtain ${\rm [Fe/H]} = -2.59\pm0.09$, ${\rm [Ba/Fe]} = -0.87\pm0.14$, and ${\rm [Eu/H]} < -2.80$. Using a 1D LTE analysis, we find $\fodd=0.02\pm0.06$, corresponding to a \element{Ba} isotopic fraction which indicates a 100\% contribution by the \textit{s}-process. This result contradicts the theory put forward by \citet{Truran81}. The result published by \citet{Lambert02} has error bars which are too broad to allow one to state conclusively that HD\,140283 is \textit{r}-process dominated. We have set a new lower limit to the [Ba/Eu] ratio, ${\rm [Ba/Eu]} > -0.66$. This lower limit marginally rules out a pure \textit{r}-process ratio in HD\,140283, consistent with the isotopic fraction for barium. A new high resolution spectrum with a greater $S/N$ around the 4129\,\AA\ line is needed to constrain a genuine abundance for \element{Eu}.

We have also carried out a careful examination of the 4934\,\AA\ line, which is more sensitive to the effects of hyperfine-splitting. We found that, due to the lack of laboratory $gf$ data surrounding the \element{Fe} blend affecting the wings of this line, it is less effective as a tool to analyse the isotopic mixture than \element{Ba} 4554\,\AA.

By examining the spectral residuals for 93 \element{Fe} lines and for \element{Ba} 4554, 4934\,\AA, we find strong line asymmetries in the red wing. These may show the shortcomings of using a 1D LTE analysis to explore isotope ratios; using a more sophisticated 3D analysis may be warranted. We are looking to take this work further in the future and analyse HD\,140283 using a 3D code. We note that \citet{Remo09} find a lower \textit{r}-process fraction for HD\,140283 using a 3D analysis than they find for 1D. If our findings are similar, then this will further accentuate the difference between the analysis of HD\,140283 and the expectations based on Truran's hypothesis. 

We set a new limit on the rotation of HD\,140283: $v\sin{i}<3.9\kms$.

\begin{acknowledgements} The authors would like to acknowledge Satoshi Kawanomoto for reducing the stellar spectrum used in this work. AJG and SGR would like to thank Gillian Nave and Juliet Pickering for kindly investigating \element{Fe} spectra to constrain spectroscopic information about a known \element{Fe} blend in the \element{Ba} 4934\,\AA\ line which was crucial for the present investigation. SGR wishes to acknowledge PhD thesis work by \citet{Blake04} on a spectrum of HD\,140283 which, while not having the benefit of a large number of \element{Fe} lines for the determination of macroturbulence, nevertheless gave a value $\fodd=0.08\pm0.11$. \end{acknowledgements}

\nocite{*}
\bibliographystyle{aa}
\bibliography{reference}

\clearpage 
\onecolumn 
\appendix
\section{Online Material}
{\bf Table A.1.} The list of iron lines used to constrain $\nu_{\rm conv}$ including measured $W$ and $\nu_{\rm obs}$.	
\label{sec:linelist}
\renewcommand{\thefootnote}{\alph{footnote}}																																																																			
\setcounter{footnote}{0}
\begin{longtable}{c	c	c	c	c	c	c	c	c	c}
\hline\hline																																																																			
\multicolumn{5}{c}	{Measured	data	from	observed	spectrum}	&	&	\multicolumn{4}{c}	{Results	from	$\chi^2$	code}	\\																																									\cline{1-5}	\cline{7-10}																																																																		
Wavelength\footnote{Lines marked with an asterisk denote those excluded from the lower panel of Fig. \ref{fig:fecomp} due to contamination with other lines found $\pm0.3$\,\AA\ from the \element{Fe} line centre.}	(\AA)	&	$W$\footnote{Equivalent	widths	of	\ion{Fe}{i}	and	\ion{Fe}{ii}	lines	measured	in	this	work.}	(m\AA)	&	$W$\footnote{Equivalent	widths	of	\ion{Fe}{i}	and	\ion{Fe}{ii}	lines	measured	by	\citet{Hosford09}	and	\citet{Lambert02}.}	(m\AA)	&	$\nu_{\rm	obs}	\ (\kms)$	&	Ion	&	&	$\Gamma	\	(\kms)$	&	\textit{A}(Fe)\footnote{$A({\rm	X})	\equiv	{\rm	log}_{10}\big(\frac{N({\rm	X})}{N({\rm	H})}\big)	+	12$}	&	$\Delta	\lambda$	(m\AA)	&	$\chi^{2}_r$	\\
\hline																																																																			
4118.54*	&	28.8	&	32.2	&	6.97	&	\ion{Fe}{i}	&	&	5.61	&	4.83	&	-8.2	&	3.3	\\																																																	
4132.90	&	14.6	&	14.6	&	6.59	&	\ion{Fe}{i}	&	&	5.85	&	4.99	&	-9.6	&	4.6	\\																																																	
4134.68	&	26.9	&	28.1	&	6.84	&	\ion{Fe}{i}	&	&	5.42	&	4.91	&	-8.8	&	1.6	\\																																																	
4137.00	&	11.1	&	11.7	&	7.07	&	\ion{Fe}{i}	&	&	5.58	&	4.79	&	-6.9	&	0.9	\\																																																	
4143.41	&	34.2	&	33.7	&	7.25	&	\ion{Fe}{i}	&	&	5.61	&	4.83	&	-8.3	&	5.4	\\																																																	
4147.67	&	22.8	&	23.5	&	6.61	&	\ion{Fe}{i}	&	&	5.52	&	4.92	&	-11.4	&	1.9	\\																																																	
4153.90	&	18.2	&	19.2	&	6.96	&	\ion{Fe}{i}	&	&	5.57	&	4.91	&	-8.5	&	0.7	\\																																																	
4154.50	&	21.0	&	22.2	&	6.65	&	\ion{Fe}{i}	&	&	5.65	&	4.84	&	-8.6	&	2.1	\\																																																	
4154.81	&	16.1	&	15.9	&	6.65	&	\ion{Fe}{i}	&	&	5.40	&	4.90	&	-9.1	&	0.9	\\																																																	
4156.80*	&	22.7	&	24.6	&	7.18	&	\ion{Fe}{i}	&	&	6.00	&	4.98	&	-8.9	&	5.3	\\																																																	
4157.78	&	13.7	&	13.2	&	6.71	&	\ion{Fe}{i}	&	&	5.86	&	4.90	&	-10.7	&	1.4	\\																																																	
4174.91*	&	14.3	&	14.7	&	6.58	&	\ion{Fe}{i}	&	&	5.58	&	4.96	&	-11.5	&	8.7	\\																																																	
4175.64	&	20.3	&	19.9	&	6.63	&	\ion{Fe}{i}	&	&	5.65	&	4.97	&	-8.9	&	4.4	\\																																																	
4176.57	&	11.6	&	11.9	&	6.90	&	\ion{Fe}{i}	&	&	5.90	&	5.06	&	-9.6	&	2.5	\\																																																	
4178.86	&	18.5	&	...	&	6.86	&	\ion{Fe}{ii}	&	&	5.82	&	4.95	&	-4.7	&	3.2	\\																																																	
4181.76*	&	38.8	&	39.7	&	7.13	&	\ion{Fe}{i}	&	&	5.80	&	4.91	&	-9.2	&	3.4	\\																																																	
4184.89	&	17.1	&	17.1	&	6.86	&	\ion{Fe}{i}	&	&	5.80	&	4.89	&	-10.5	&	1.8	\\																																																	
4187.04	&	46.8	&	46.4	&	7.22	&	\ion{Fe}{i}	&	&	5.79	&	4.84	&	-11.1	&	4.7	\\																																																	
4187.80*	&	49.5	&	50.2	&	7.46	&	\ion{Fe}{i}	&	&	6.07	&	4.87	&	-12.1	&	11.7	\\																																																	
4191.43	&	39.5	&	38.8	&	7.04	&	\ion{Fe}{i}	&	&	5.93	&	4.87	&	-7.8	&	7.2	\\																																																	
4199.09	&	48.7	&	49.0	&	7.32	&	\ion{Fe}{i}	&	&	5.86	&	4.80	&	-8.7	&	14.1	\\																																																	
4210.34	&	31.9	&	32.2	&	7.40	&	\ion{Fe}{i}	&	&	5.84	&	4.89	&	-9.8	&	2.0	\\																																																	
4217.54	&	12.0	&	13.9	&	7.13	&	\ion{Fe}{i}	&	&	5.46	&	4.85	&	-9.5	&	1.2	\\																																																	
4219.36	&	23.5	&	23.5	&	7.25	&	\ion{Fe}{i}	&	&	6.04	&	4.92	&	-8.6	&	2.4	\\																																																	
4222.21	&	29.0	&	28.5	&	6.90	&	\ion{Fe}{i}	&	&	5.88	&	4.88	&	-8.9	&	6.8	\\																																																	
4225.45	&	13.2	&	14.4	&	6.70	&	\ion{Fe}{i}	&	&	5.52	&	4.95	&	-7.7	&	0.9	\\																																																	
4233.17*	&	43.0	&	43.2	&	7.09	&	\ion{Fe}{ii}	&	&	5.97	&	4.87	&	-3.8	&	9.5	\\																																																	
4233.60	&	43.4	&	43.7	&	7.10	&	\ion{Fe}{i}	&	&	5.81	&	4.87	&	-10.0	&	7.2	\\																																																	
4238.81	&	20.4	&	21.6	&	7.04	&	\ion{Fe}{i}	&	&	6.00	&	4.91	&	-9.3	&	6.5	\\																																																	
4282.40	&	46.9	&	45.7	&	7.12	&	\ion{Fe}{i}	&	&	5.74	&	4.84	&	-8.8	&	10.0	\\																																																	
4337.05	&	36.1	&	41.9	&	7.01	&	\ion{Fe}{i}	&	&	5.71	&	4.90	&	-14.3	&	3.9	\\																																																	
4352.73*	&	24.3	&	24.4	&	6.77	&	\ion{Fe}{i}	&	&	5.88	&	4.91	&	-10.2	&	22.3	\\																																																	
4416.82	&	10.9	&	11.8	&	6.43	&	\ion{Fe}{ii}	&	&	5.42	&	4.97	&	-9.1	&	0.4	\\																																																	
4430.61	&	13.5	&	13.2	&	6.73	&	\ion{Fe}{i}	&	&	5.77	&	5.00	&	-9.7	&	1.8	\\																																																	
4442.34	&	30.3	&	31.2	&	7.17	&	\ion{Fe}{i}	&	&	5.88	&	4.92	&	-9.8	&	4.1	\\																																																	
4443.19	&	11.7	&	9.9	&	6.64	&	\ion{Fe}{i}	&	&	5.56	&	4.85	&	-9.6	&	0.3	\\																																																	
4447.72	&	24.5	&	24.0	&	6.90	&	\ion{Fe}{i}	&	&	5.85	&	4.94	&	-10.9	&	4.2	\\																																																	
4461.65	&	40.9	&	42.4	&	6.69	&	\ion{Fe}{i}	&	&	5.39	&	5.01	&	-13.7	&	6.4	\\																																																	
4466.55	&	29.4	&	30.3	&	7.01	&	\ion{Fe}{i}	&	&	5.80	&	4.88	&	-12.7	&	3.0	\\																																																	
4489.74	&	11.7	&	11.6	&	6.76	&	\ion{Fe}{i}	&	&	5.56	&	4.97	&	-14.9	&	1.3	\\																																																	
4494.56	&	34.4	&	34.6	&	6.95	&	\ion{Fe}{i}	&	&	5.73	&	4.94	&	-13.1	&	4.3	\\																																																	
4508.28	&	16.2	&	17.2	&	6.88	&	\ion{Fe}{ii}	&	&	5.97	&	5.01	&	-10.2	&	4.7	\\																																																	
4515.33	&	11.8	&	13.5	&	6.61	&	\ion{Fe}{ii}	&	&	5.79	&	4.96	&	-6.0	&	1.2	\\																																																	
4520.22	&	11.2	&	12.8	&	6.63	&	\ion{Fe}{ii}	&	&	5.68	&	4.98	&	-10.9	&	1.3	\\																																																	
4522.63*	&	23.3	&	24.7	&	7.01	&	\ion{Fe}{ii}	&	&	6.12	&	4.88	&	-10.8	&	52.4	\\																																																	
4531.15	&	22.3	&	22.6	&	6.70	&	\ion{Fe}{i}	&	&	5.54	&	4.90	&	-10.6	&	0.7	\\																																																	
4555.89*	&	15.8	&	18.3	&	6.76	&	\ion{Fe}{ii}	&	&	5.81	&	4.89	&	-12.6	&	4.3	\\																																																	
4583.83	&	37.4	&	38.6	&	6.98	&	\ion{Fe}{ii}	&	&	5.76	&	5.13	&	-11.3	&	2.6	\\																																																	
4602.94	&	21.6	&	21.9	&	6.97	&	\ion{Fe}{i}	&	&	5.75	&	4.97	&	-11.0	&	2.9	\\																																																	
4736.77	&	14.0	&	14.6	&	6.89	&	\ion{Fe}{i}	&	&	5.78	&	4.99	&	-14.2	&	1.7	\\																																																	
4871.33	&	37.1	&	...	&	7.25	&	\ion{Fe}{i}	&	&	5.93	&	4.84	&	-12.2	&	5.5	\\																																																	
4872.14	&	27.8	&	...	&	7.12	&	\ion{Fe}{i}	&	&	5.93	&	4.87	&	-11.6	&	10.9	\\																																																	
4890.76	&	35.9	&	...	&	7.09	&	\ion{Fe}{i}	&	&	5.81	&	4.86	&	-15.1	&	10.3	\\																																																	
4918.99	&	38.8	&	...	&	7.11	&	\ion{Fe}{i}	&	&	5.85	&	4.88	&	-11.0	&	15.3	\\																																																	
4938.81	&	12.1	&	...	&	6.80	&	\ion{Fe}{i}	&	&	5.59	&	4.89	&	-15.0	&	2.3	\\																																																	
4994.13	&	13.1	&	14.0	&	6.50	&	\ion{Fe}{i}	&	&	5.58	&	4.88	&	-12.4	&	0.8	\\																																																	
5001.86	&	13.3	&	...	&	7.01	&	\ion{Fe}{i}	&	&	5.91	&	4.87	&	-11.1	&	2.2	\\																																																	
5006.12	&	27.5	&	...	&	6.98	&	\ion{Fe}{i}	&	&	5.68	&	4.84	&	-13.1	&	1.4	\\																																																	
5012.07	&	32.8	&	32.0	&	6.78	&	\ion{Fe}{i}	&	&	5.58	&	4.98	&	-17.9	&	8.8	\\																																																	
5041.07	&	13.2	&	...	&	6.89	&	\ion{Fe}{i}	&	&	6.15	&	5.07	&	-12.5	&	19.1	\\																																																	
5049.82	&	22.6	&	...	&	6.90	&	\ion{Fe}{i}	&	&	5.80	&	4.92	&	-15.6	&	3.6	\\																																																	
5051.64	&	22.8	&	...	&	6.61	&	\ion{Fe}{i}	&	&	5.60	&	4.97	&	-15.6	&	5.4	\\																																																	
5068.77	&	11.2	&	...	&	6.76	&	\ion{Fe}{i}	&	&	5.33	&	4.85	&	-17.4	&	0.7	\\																																																	
5083.34	&	15.3	&	16.0	&	6.46	&	\ion{Fe}{i}	&	&	5.56	&	4.88	&	-14.6	&	3.8	\\																																																	
5098.70	&	10.2	&	...	&	7.34	&	\ion{Fe}{i}	&	&	5.52	&	5.03	&	-15.9	&	0.6	\\																																																	
5107.45*	&	11.6	&	12.0	&	6.44	&	\ion{Fe}{i}	&	&	5.51	&	5.01	&	-20.3	&	5.1	\\																																																	
5110.41	&	24.1	&	25.0	&	6.61	&	\ion{Fe}{i}	&	&	5.54	&	5.06	&	-18.8	&	4.3	\\																																																	
5123.72	&	11.0	&	12.0	&	6.40	&	\ion{Fe}{i}	&	&	5.61	&	4.99	&	-15.1	&	2.7	\\																																																	
5142.93*	&	11.8	&	...	&	7.25	&	\ion{Fe}{i}	&	&	6.20	&	4.95	&	-14.5	&	46.7	\\																																																	
5162.27	&	12.6	&	...	&	7.25	&	\ion{Fe}{i}	&	&	5.76	&	5.06	&	-10.3	&	0.6	\\																																																	
5166.28	&	10.3	&	...	&	6.33	&	\ion{Fe}{i}	&	&	5.41	&	4.97	&	-15.4	&	0.9	\\																																																	
5171.60	&	41.2	&	...	&	6.86	&	\ion{Fe}{i}	&	&	5.44	&	4.67	&	0.3	&	18.0	\\																																																	
5191.46	&	21.9	&	...	&	6.86	&	\ion{Fe}{i}	&	&	5.81	&	4.86	&	-16.5	&	10.5	\\																																																	
5192.34	&	28.4	&	...	&	6.85	&	\ion{Fe}{i}	&	&	5.78	&	4.84	&	-16.9	&	29.3	\\																																																	
5194.94	&	24.4	&	26.0	&	6.67	&	\ion{Fe}{i}	&	&	5.79	&	4.93	&	-19.2	&	7.2	\\																																																	
5216.27	&	19.6	&	...	&	6.65	&	\ion{Fe}{i}	&	&	5.60	&	4.89	&	-14.9	&	4.7	\\																																																	
5234.64	&	12.9	&	13.0	&	6.57	&	\ion{Fe}{ii}	&	&	5.60	&	4.89	&	-15.8	&	1.9	\\																																																	
5266.56	&	32.9	&	...	&	7.09	&	\ion{Fe}{i}	&	&	5.87	&	4.85	&	-14.0	&	8.6	\\																																																	
5281.79	&	13.5	&	...	&	6.81	&	\ion{Fe}{i}	&	&	5.77	&	4.86	&	-17.4	&	5.1	\\																																																	
5283.62	&	19.5	&	...	&	7.02	&	\ion{Fe}{i}	&	&	5.89	&	4.94	&	-17.1	&	12.1	\\																																																	
5302.31	&	10.2	&	...	&	6.85	&	\ion{Fe}{i}	&	&	5.93	&	4.86	&	-13.8	&	2.8	\\																																																	
5324.19	&	35.4	&	...	&	7.15	&	\ion{Fe}{i}	&	&	5.91	&	5.10	&	-15.2	&	21.5	\\																																																	
5339.93	&	12.7	&	...	&	6.90	&	\ion{Fe}{i}	&	&	5.88	&	4.95	&	-18.5	&	4.7	\\																																																	
5367.47	&	11.3	&	...	&	7.04	&	\ion{Fe}{i}	&	&	5.81	&	4.84	&	-12.5	&	5.0	\\																																																	
5369.96	&	14.1	&	...	&	7.11	&	\ion{Fe}{i}	&	&	5.75	&	4.81	&	-13.0	&	5.1	\\																																																	
5383.37	&	18.4	&	...	&	7.01	&	\ion{Fe}{i}	&	&	5.65	&	4.78	&	-15.3	&	10.0	\\																																																	
5393.17	&	12.2	&	...	&	7.20	&	\ion{Fe}{i}	&	&	6.08	&	4.88	&	-14.2	&	4.1	\\																																																	
5404.15	&	18.4	&	...	&	7.52	&	\ion{Fe}{i}	&	&	5.65	&	4.93	&	-4.3	&	4.6	\\																																																	
5415.20	&	17.5	&	...	&	7.26	&	\ion{Fe}{i}	&	&	5.80	&	5.28	&	-13.6	&	2.3	\\																																																	
5569.62	&	12.9	&	...	&	7.19	&	\ion{Fe}{i}	&	&	5.97	&	4.84	&	-15.2	&	4.2	\\																																																	
5572.85	&	19.7	&	...	&	7.05	&	\ion{Fe}{i}	&	&	5.94	&	4.29	&	-16.0	&	4.1	\\																																																	
5615.66	&	34.1	&	...	&	7.18	&	\ion{Fe}{i}	&	&	5.96	&	4.80	&	-15.4	&	5.6	\\																																																	
6252.56	&	11.3	&	14.0	&	6.74	&	\ion{Fe}{i}	&	&	5.79	&	5.00	&	-21.1	&	2.6	\\																																																	
\hline																																																																			
\label{tab:Felines}
\end{longtable}																																																																			
\renewcommand{\thefootnote}{\arabic{footnote}}																																																																			

\end{document}